\documentclass[superscriptaddress,aps,preprintnumbers,amsmath,amssymb,nofootinbib,10pt,onecolumn,prd,floatfix]{revtex4-2}
\usepackage{physics, amsfonts, tensor, mathrsfs, mathtools, paralist, comment, setspace, enumerate, enumitem, xcolor}
\usepackage[font=small,labelfont=bf,justification=justified, singlelinecheck = false]{subcaption}
\usepackage[font=small,labelfont=bf,justification=raggedright, singlelinecheck = false]{caption}
\usepackage{graphicx}
\usepackage{hyperref}
\hypersetup{%
bookmarksnumbered=true,%
bookmarksopen=true,%
colorlinks=true,%
linkcolor=blue,
citecolor=red,
}

\newcommand{\figref}[1]{FIG.~\ref{#1}}
\newcommand{\secref}[1]{Sec.~\ref{#1}}
\newcommand{\appref}[1]{Appendix~\ref{#1}}

\begin{document}
\count\footins = 1000
\preprint{YITP-24-152, IPMU24-0040}

\title{Boulware Vacuum vs. Regularity: \\
Thoughts on Anomaly-Induced Effective Action
}

\author{Kota Numajiri}%
\email{numajiri.kota.m3@s.mail.nagoya-u.ac.jp}
\affiliation{
Department of Physics, Nagoya University, Nagoya 464-8602, Japan}

\author{Kazumasa Okabayashi}%
\email{kazumasa.okabayashi@yukawa.kyoto-u.ac.jp}
\affiliation{
Center for Gravitational Physics and Quantum Information,
Yukawa Institute for Theoretical Physics, Kyoto University, 606-8502, Kyoto, Japan
}

\author{Shinji Mukohyama}%
\email{shinji.mukohyama@yukawa.kyoto-u.ac.jp}
\affiliation{
Center for Gravitational Physics and Quantum Information,
Yukawa Institute for Theoretical Physics, Kyoto University, 606-8502, Kyoto, Japan
}
\affiliation{Kavli Institute for the Physics and Mathematics of the Universe (WPI),
The University of Tokyo Institutes for Advanced Study,
The University of Tokyo, Kashiwa, Chiba 277-8583, Japan}

\begin{abstract}

We examine the vacuum state and its corresponding renormalized stress-energy tensor (RSET) in static horizonless regular spacetime in both two and four dimensions. Using the local field formulation of the anomaly-induced effective action, we show that the regularities of the spacetime and the RSET dictate the appropriate vacuum state. Furthermore, through a case study under the horizonless Bardeen-type spacetime, we demonstrate that the preferred vacuum state is not the Boulware vacuum, but a nontrivial one with a different RSET profile.

\end{abstract}

\maketitle

\section{Introduction}

In curved spacetime, quantum fields exhibit an interesting behavior known as the trace (or conformal) anomaly \cite{Capper:1974ic, Capper:1975ig, Deser:1976yx, Duff:1977ay} (see also \cite{Birrell:1982ix, Parker:2009uva, Hu:2020luk, Duff_1994} for comprehensive review). A curved geometry generally induces gravitational vacuum polarization \cite{Hawking:1974rv, Davies:1974th, Hawking:1975vcx, Wald:1975kc, Parker:1975jm} and/or dynamical particle production \cite{Parker:1968mv, Parker:1969au, Sexl:1969ix, Zeldovich:1970si}. Such effects modify the expectation value of the quantum field's stress-energy tensor (SET) operator, which is described as the regularization and renormalization procedures \cite{Wald:1978ce, 1977RSPSA.354...59D, Fulling:1977jm}. Consequently, the renormalized stress-energy tensor (RSET) and its trace have a nontrivial expectation value. Even a conformally coupled massless field, whose SET is traceless at the classical level, acquires a nonzero anomalous trace value. This trace anomaly of RSET is first discussed in \cite{Capper:1974ic, Capper:1975ig, Deser:1976yx, Duff:1977ay}, and then applied to the case of a massless field in the black hole (BH) spacetime \cite{Christensen:1977jc, Candelas:1980zt, Frolov:1987gw, Anderson:1994hg} to study Hawking radiation \cite{Hawking:1974rv, Hawking:1975vcx}. In the meantime, the cosmological spacetime case has been discussed \cite{Bunch:1977sc, Davies:1977ze, Page:1982fm, Brown:1986jy}, and extended to more general spacetimes \cite{Parker:1978gh}, as well as the investigation of the backreaction on the spacetime (semiclassical gravity) \cite{1974JETP...39..742L, Grishchuk:1977zz, Guth:1980zm, HAJICEK19809, Bardeen:1981zz}.

The value of trace anomaly is determined by the local geometry, which is also responsible for regularizing the UV divergence of SET. On the other hand, the value of each component of RSET significantly depends also on the quantum state. The vacuum state of the quantum field in curved spacetime must be specified on some Cauchy surfaces. The choice of vacuum is closely related to the boundary conditions at the spacetime boundaries, as we will demonstrate explicitly later. Therefore, the choice of vacuum exhibits global properties of the spacetime.

In the case of the four-dimensional Schwarzschild spacetime, there are three well-known vacuum choices:
\begin{inparaenum}[(I)]
    \item the Israel-Hartle-Hawking vacuum \cite{Israel:1976ur, Hartle:1976tp}, which describes a BH in thermal equilibrium,
    \item the Unruh vacuum \cite{Unruh:1976db}, which corresponds to an evaporating or collapsing BH with nonzero flux,
    \item the Boulware vacuum \cite{Boulware:1974dm, Boulware:1975fe}, which coincides with the Minkowski vacuum as $r \rightarrow \infty$.
\end{inparaenum}
Among these choices, the Boulware vacuum aligns with our intuition of the vacuum for observers at a greater distance from the black hole. However, it is well known that this vacuum leads to a divergent RSET near the horizon.

In this study, we extend the analysis to the case of horizonless regular spacetimes with spherical symmetry. One key difference from BH cases is the regularity of the spacetime. This requires the quantum fields and RSET to be regular everywhere in the bulk of the spacetime. Another difference arises from the global structure of the spacetime. Spacetimes with black hole horizons are not isomorphic to those without horizons, leading to different possible vacuum choices. As a result, based on the local field formulation for the anomaly-induced effective action explained later, we demonstrate that the Boulware vacuum is not consistent with the central regularities in four-dimensional horizonless spacetime through a case study of a horizonless Bardeen-type spacetime \cite{1968qtr..conf...87B, Carballo-Rubio:2022nuj}, while it is consistent in the two-dimensional case.

There are some previous discussions in \cite{Carballo-Rubio:2017tlh, Arrechea:2021xkp, Reyes:2023fde, Arrechea:2023oax} which focused on the effect of massless fields on the geometry of horizonless compact stars such as a uniform density star. One of the main findings was that the semiclassical effect leads to exceeding the Buchdahl limit, which is the maximum allowable compactness of a star. All of the aforementioned works assumed the Boulware vacuum. However, it should be noted that this assumption of the Boulware vacuum in a regular spacetime is nontrivial. In particular, if low-frequency ($\ll 1/(\mbox{radius})$) mode functions of quantum fields contribute non-negligibly to their RSET in the distant region, then the asymptotic behavior of the RSET with the Killing vacuum may exhibit nontrivial dependence on the internal structure of the horizonless compact object.
\footnote{
In the context of tidal deformation analysis in compact objects, it is widely known that the asymptotic behavior of perturbations, or mode functions, in the distant region is significantly influenced by the internal structure of the compact object. For low-frequency mode functions, this leads to a nontrivial (static and dynamical) tidal Love number that depends on the internal structure of the regular
compact object~\cite{Flanagan:2007ix, Damour:2009vw,
Binnington:2009bb, Poisson:2020vap, HegadeKR:2024agt}.} 
This suggests that the exact matching with the Boulware RSET might require some fine-tuning of the internal structure. Our work sheds light on the validity of assuming the Boulware vacuum in a regular spacetime, illustrating the relationship between the central regularity of RSET and the choice of vacuum.

To easily and systematically investigate the relation among the spacetime structure, vacuum state, and the RSET, we adopted the local auxiliary field (``conformalon'') formulation \cite{Riegert:1984kt, Shapiro:1994ww, Balbinot:1999ri, Balbinot:1999vg, Mazur:2001aa, Mottola:2006ew, Barcelo:2011bb, Shen:2015zya} of the trace anomaly. Typically, each regularization method heavily relies on the background spacetime, making it challenging to extend the investigation to arbitrary spacetime scenarios, including those without horizons. However, by using the local field formulation, the analysis of quantum effects can be simplified to solving the profiles of local fields with boundary conditions that correspond to different vacuum choices. Although the application for nonconformally flat spacetime cases gives not exact but just approximation results, this formulation makes it much easier to extend the study to arbitrary geometries and provides a clearer understanding of the possible vacuum states.

In this paper, we organize the content as follows: First, we start with a discussion of the 2D case in \secref{sec:2D}, providing a brief introduction to the local auxiliary field formulation in \secref{sec:2D_general}. Subsequently, after a review of the two-dimensional Minkowski and Schwarzschild spacetime cases, we investigate the possible vacuum and the resulting RSET in a horizonless static spacetime in \secref{sec:2D_sss}. Then we move on to the four-dimensional case in \secref{sec:4D}. We introduce the general formalism for four-dimensional cases in \secref{sec:4D_general}, review the results in the Schwarzschild case in \secref{sec:4D_Sch}. Afterward, we analyze the case of a horizonless static spacetime with spherical symmetry through a case study in a Bardeen-type scenario in \secref{sec:4D_horizonless}. 
Finally, \secref{sec:discussion} is dedicated to discussion and conclusion. Appendix explains the junction condition of the local fields concerning their application to general compact star geometries.

As conventions, the speed of light $c$ and the Newton constant $G_N$ are taken as unity. We use $\qty{-1,1,1, \cdots, 1}$ as the metric signs, and other curvature conventions are based on \cite{Carroll:2004st}.

\section{Two-dimensional case \label{sec:2D}}
\subsection{The local action for trace anomaly \label{sec:2D_general}}

Based on \cite{Mottola:2006ew}, we begin by reviewing the trace anomaly and corresponding local action in two-dimensional spacetime. The trace anomaly in two-dimensional spacetime is expressed as \cite{Birrell:1982ix}
\begin{align}
    \ev{ T^a_a }=\frac{N}{24 \pi} R.
    \label{eq:Anom2d}
\end{align}
The corresponding anomalous effective action that reproduces this trace anomaly can be written as \cite{Riegert:1984kt, Mazur:2001aa}
\begin{align}
    S_{\mathrm{anom}}^{(2)}[g]
    =\frac{Q^2}{16 \pi} 
    \int \dd[2]{x} \sqrt{-g} \int \dd[2]{x^{\prime}} 
    \sqrt{-g^{\prime}} 
    R(x) \square^{-1}\qty(x, x^{\prime}) R(x^{\prime}) ,
    \label{eq:NLaction2d}
\end{align}
where $Q^2 = - N /6$ for the case with $N$ massless matter fields.\footnote{In the case with the metric perturbations besides matter fields, the number $N$ should be replaced with $N-25$ \cite{Knizhnik:1988ak, David:1988hj, Distler:1988jt}.} 
The term $\square^{-1}\qty(x, x^{\prime})$ denotes Green's function corresponding to the inverse of the scalar d'Alembertian operator, which indicates the nonlocal nature of the RSET in addition to the dependence on the local curvature of the spacetime.

Introducing auxiliary scalar field $\varphi$ which satisfies 
\begin{align}
    \Box \varphi = -R,
    \label{eq:EOM_phi2d}
\end{align}
the nonlocal action~\eqref{eq:NLaction2d} can be represented with a local Lagrangian \cite{Mottola:2006ew}
\begin{align}
    S_{\mathrm{anom}}^{(2)} [g ; \varphi] 
    \equiv \frac{Q^2}{16 \pi} \int \dd[2]{x} \sqrt{-g}
    \qty(\nabla_a \varphi \nabla^a \varphi-2 R \varphi),
    \label{eq:Laction2d}
\end{align}
that can also be derived directly as the Wess-Zumino effective action corresponding to Eq.~\eqref{eq:Anom2d} \cite{Mazur:2001aa, Mottola:2006ew}. The variation of this action with $\varphi$ yields Eq.~\eqref{eq:EOM_phi2d} as the field equation. The nonlocal anomalous action \eqref{eq:NLaction2d} is reproduced by substituting a formal solution $\varphi = -\square^{-1} R$ for Eq.~\eqref{eq:EOM_phi2d} into Eq.~\eqref{eq:Laction2d}. 
The nonlocal feature of the RSET besides the effect of local geometry is now implemented in the solution $\varphi$ for Eq.~\eqref{eq:EOM_phi2d} with the arbitrariness of the homogeneous part.

The renormalized stress-energy tensor for $\varphi$ is obtained by the variation of Eq.~\eqref{eq:Laction2d} with the metric $g$
\begin{align}
    T_{a b}^{(2)}[g ; \varphi] 
    &\equiv
    -\frac{2}{\sqrt{-g}} \fdv{S_{\mathrm{anom}}^{(2)}[g ; \varphi]}{g^{a b}} \nonumber \\
    &= \frac{Q^2}{4\pi}
    \qty[
        -\nabla_a \nabla_b \varphi+g_{a b} \square \varphi-\frac{1}{2}\left(\nabla_a \varphi\right)\left(\nabla_b \varphi\right)
        +\frac{1}{4} g_{a b}\left(\nabla_c \varphi\right)\left(\nabla^c \varphi\right)
    ].
    \label{eq:SET2d}
\end{align}
This RSET is shown to be covariantly conserved by Eq.~\eqref{eq:EOM_phi2d}. The trace of this RSET matches the trace anomaly~\eqref{eq:Anom2d}. Therefore, Eq.~\eqref{eq:Laction2d} adequately represents the anomaly in two-dimensional spacetime. The complete low-energy effective action for two-dimensional gravity with consistent matter coupling consists of this anomalous action and other mass dimension two terms, including the Einstein-Hilbert term.

The auxiliary scalar field $\varphi$ represents the parametrizing degree of freedom for the local Weyl transformation $g_{ab} \rightarrow \mathrm{e}^{-\varphi} g_{ab}$. As a result of the field equation \eqref{eq:EOM_phi2d}, it is demonstrated that the curvature of the transformed geometry vanishes. This transformation has a degeneracy due to the degree of freedom to add a homogeneous solution of Eq.~\eqref{eq:EOM_phi2d} to $\varphi$. This lack of uniqueness relates to the additional flexibility of the boundary conditions, i.e. the choice of vacuum, which is influenced by the spacetime global properties and is essentially nonlocal, not expressed as local curvature terms.

One can notice that the field equation \eqref{eq:EOM_phi2d} for the auxiliary scalar $\varphi$ and the RSET~\eqref{eq:SET2d} is invariant under a constant shift $\varphi \rightarrow \bar{\varphi} = \varphi + \varphi_0$. This is because the Einstein-Hilbert action becomes a topological invariant in two-dimensional spacetime,
\begin{align}
    \chi=\frac{1}{4 \pi} \int \dd[2]{x} \sqrt{-g} R.
\end{align}
Thus one can introduce a topological current $\Omega^a$,
\begin{align}
    R = \nabla_a \Omega^a.
    \label{eq:Omega_2D}
\end{align}
With this current $\Omega^a$, the anomalous action~\eqref{eq:Laction2d} is rewritten as
\begin{align}
    S_{\mathrm{anom}}^{(2)} [g ; \varphi] 
    = \frac{Q^2}{16 \pi} \int_V \dd[2]{x} \sqrt{-g}
    \qty(\nabla^a \varphi+2 \Omega^a) \nabla_a \varphi
    - \frac{Q^2}{8 \pi} \oint_{\partial V} \dd{s} \sqrt{-g} \, \varphi \Omega^a n_a ,
\end{align}
The shift invariance with $\varphi$ corresponding to the global Weyl invariance implies the Noether current 
\begin{align}
    J^a \equiv \nabla^a \varphi + \Omega^a,
    \label{eq:current_2D}
\end{align}
which is covariantly conserved as is shown from Eqs.~\eqref{eq:EOM_phi2d} and \eqref{eq:Omega_2D}. The existence of such conserved current and the global charge implies the auxiliary field $\varphi$ contains information about the spacetime global properties, in addition to macroscopic quantum effects including the choice of the vacuum state.

\subsection{Stress-energy tensor and vacuum state in static spacetime \label{sec:2D_sss}}
A solution $\varphi$ of Eq.~\eqref{eq:EOM_phi2d} in a given spacetime is specified by a boundary condition, which contains information about the corresponding quantum state. Let us consider some examples to understand this feature more concretely.

\subsubsection{Flat spacetime}
First we review the case with flat spacetime \cite{Mottola:2006ew}. In the Rindler coordinate
\begin{align}
    ds^2
    =- \dd{t}^2+ \dd{x}^2
    =-\rho^2 \dd{\eta}^2 + \dd{\rho}^2,
\end{align}
the boost-invariant solution $\varphi=\varphi(\rho)$ for the field equation \eqref{eq:EOM_phi2d} is
\begin{align}
    \varphi (\rho) =2 q \log \qty(\frac{\rho}{\rho_0}),
    \label{eq:phi_flat2d}
\end{align}
with integration constants $q$ and $\rho_0$. This solution has singularities at $\rho \rightarrow 0$ and $\rho \rightarrow \infty$ for $q \neq 0$, and corresponds to the static potential distribution with a point charge of magnitude $-q$ at the origin and $+q$ at infinity in Euclidean space. The contribution at $\rho=0$ can be picked up via
\begin{align}
    q =\frac{1}{2} \int_{\partial \Sigma} J^a \dd{\Sigma}_a ,
\end{align}
where $\partial\Sigma$ is the boundary of the spacelike Cauchy surface $\Sigma$. This implies that in general nonzero $q$ means a kind of topological defect located at the Rindler horizon ($\rho=0$). 

The RSET is obtained by substituting Eq.~\eqref{eq:phi_flat2d} into Eq.~\eqref{eq:SET2d}
\begin{align}
    T\indices{_a^b}
    =\frac{Q^2}{4 \pi} \frac{(2-q) q}{\rho^2}
    \mqty(
        -1 & 0 \\
        0 & 1
    ).
\end{align}
Both the trivial solution with $q=0$ representing the Minkowski vacuum and the nontrivial solution with $q=2$ lead to vanishing RSET. Indeed, the rescaled metrics $\mathrm{e}^{-\varphi}ds^2$ with these solutions are equivalent through the change of variables $\rho \rightarrow \rho_0^2 / \rho$. These facts indicate the existence of the topologically twofold degenerate vacuums in two-dimensional flat space. The choice $q=1$ gives nonzero RSET which diverges as $\rho \rightarrow 0$ while the rescaled metric is again flat. This solution corresponds to the Fulling-Rindler state.

\subsubsection{Schwarzschild}
The general static spacetime in two dimensions can be written as
\begin{align}
    d s^2
    =-f(r) \dd{t}^2+\frac{\dd{r}^2}{g(r)}
    =-f(r) \dd{t}^2+\frac{\dd{\bar{r}}^2}{f(r)}
    =f\qty(-\dd{t}^2+\dd{r}^{* 2}).
    \quad \qty(\dd{r}^*=\frac{\dd{\bar{r}}}{f}=\frac{\dd{r}}{\sqrt{fg}})
    \label{eq:SSS_2d}
\end{align}
This class of solutions includes the Schwarzschild solution $f(r) = g(r)^{-1} = 1-2M/r$, and the (static chart of) de Sitter solution $f(r) = g(r)^{-1} = 1-H^2 r^2$.
By using the field equation \eqref{eq:EOM_phi2d}, the RSET~\eqref{eq:SET2d} in such spacetimes \cite{Mottola:2006ew} is
\begin{align}
    T\indices{_{r^*}^{r^*}}
    &=\frac{N}{24 \pi}
    \qty{
        \frac{1}{f} \pdv[2]{\varphi}{r^{*}\,}
        -\frac{f^{\prime}}{2 f} \pdv{\varphi}{r^{*}} 
        +\frac{1}{4 f}\left(\qty(\pdv{\varphi}{r^{*}})^2+\dot{\varphi}^2\right)+R
    }, \nonumber \\
    T\indices{_{t}^{t}}
    &= \frac{N}{24 \pi}
    \qty{
        -\frac{\ddot{\varphi}}{f}+\frac{f^{\prime}}{2 f} \pdv{\varphi}{r^{*}} -\frac{1}{4 f}\qty( \qty(\pdv{\varphi}{r^{*}})^2+\dot{\varphi}^2)+R
    },
    \nonumber \\
    T\indices{_{t}^{r^*}}
    &= \frac{N}{24 \pi f}
    \qty{
        \pdv{\varphi}{r^{*}}{t}-\frac{f^{\prime}}{2} \dot{\varphi}+\frac{1}{2} \dot{\varphi} \dv{\varphi}{r^{*}} 
    } .
    \label{eq:RSET_SSS_2d}
\end{align}
Primes and dots denote the derivative with respect to $r$ and $t$ respectively.
To avoid time-dependent RSET, $\varphi(t,r^*)$ must be at most linear in $t$ and the coefficient of $t$ must be constant. With this restriction, the solution for Eq.~\eqref{eq:EOM_phi2d} is found to be \cite{Mottola:2006ew}
\begin{align}
    &\varphi 
    =c_0+\frac{q}{2 M} r^*+\frac{p}{2 M} t+\ln f, 
    \label{eq:phi_2d_sss}
\end{align}
with constants $c_0, q,$ and $p$. The last inhomogeneous term introduces the local curvature effect of this spacetime. 

In the Schwarzschild spacetime with $f(r)= 1-2M/r$, the RSET~\eqref{eq:RSET_SSS_2d} reduces to \cite{Christensen:1977jc, Mottola:2006ew}
\begin{align}
    T\indices{_{r^*}^{r^*}}
    &=\frac{N}{24 \pi}
    \qty{
        -\frac{1}{4 f}\qty(\frac{p^2+q^2}{4 M^2}-\frac{4 M^2}{r^4})+\frac{4 M}{r^3}
    }, \nonumber \\
    T\indices{_{t}^{t}}
    &= \frac{N}{96 \pi f}
    \qty(\frac{p^2+q^2}{4 M^2}-\frac{4 M^2}{r^4}),
    \nonumber \\
    T\indices{_{t}^{r^*}}
    &=\frac{N}{192 \pi M^2} \frac{p q}{f}.
    \label{eq:RSET_Sch_2d}
\end{align}
The finiteness conditions of null energy flux on the future and past horizons lead to the equations $(q-p)^2 = 1$ and $(q+p)^2 = 1$, respectively. From these equations, the choice $p=q=0$ is incompatible. Thus, these finite conditions mean that the solution of $\varphi$ \eqref{eq:phi_2d_sss} must blow up as $r^* \rightarrow \infty$ and/or as $t \rightarrow \pm \infty$, unlike in flat spacetime where the trivial solution is allowed even under the finiteness condition at the Rindler horizon.

One possible state that satisfies the conditions is the Israel-Hartle-Hawking state \cite{Hartle:1976tp}, whose RSET is diagonal and reduces to the SET for perfect gas with the Hawking temperature $T_H=(8\pi M)^{-1}$ at both $r\rightarrow 2M$ and  $r\rightarrow \infty$. This state is achieved when either $p$ or $q$ is zero while the other is 1. Another example with the finiteness at the future horizon is the Unruh state \cite{Unruh:1976db} with $p=-q=1/2$, which has nonvanishing energy flux $T\indices{_{t}^{r^*}}$.

The aforementioned choice of parameters $p=q=0$ corresponds to the Boulware vacuum \cite{Boulware:1974dm, Boulware:1975fe} where $T\indices{_a^b} \rightarrow 0$ for $r \rightarrow \infty$. With this state the asymptotic form of the RSET around the horizons is
\begin{align}
    \eval{T\indices{_a^b}}_B 
    \rightarrow-\frac{\pi}{6} \frac{T_H^2}{f}
    \left(\begin{array}{cc}
        -1 & 0 \\
        0 & 1
    \end{array}\right), 
    \quad \qty(r \rightarrow 2 M)
\end{align}
which diverges as $r\rightarrow 2M$. Therefore, it is impossible to simultaneously make RSET finite at $r\rightarrow 2M$ and vanishing for $r\rightarrow \infty$ in the Schwarzschild spacetime. This results from a global topological effect of spacetime with an event horizon, which means a nonzero Euclidean Euler number.

\subsubsection{General static horizonless spacetime \label{sec:2D_SSS}}

In the case of general asymptotically Schwarzschild regular spacetime with a globally finite metric function $f$, the solution~\eqref{eq:phi_2d_sss} for $\varphi$ and the RSET~\eqref{eq:RSET_SSS_2d} still hold. Note that here we assume $Z_2$ symmetry with respect to some origin ($r=0$) as the counterpart of spherical symmetry in 4D spacetime. At $r \rightarrow \infty$, these solutions agree with those in the Schwarzschild spacetime. On the other hand, the regularity of the metric function $f$ improves the divergence of the RSET around the center regardless of the state. At $r \rightarrow 0$ the smooth regular metric function should behave as:
\begin{align*}
    f(r^*) = f_0 + \frac{1}{2} f_2 r^{*2} + \frac{1}{4!} f_4 r^{*4} + \order{r^{*6}},
    \quad \qty(r^*\rightarrow 0)
\end{align*}
with $f_0 > 0$, where we have set $r^*=0$ at $r=0$. Then the asymptotic solution $\varphi$~\eqref{eq:phi_2d_sss} should reduce to 
\begin{align}
    \varphi(r^*) = c_0 + \log f_0 + \frac{p}{2 M} t + \frac{q}{2 M} r^*
    + \frac{f_2}{2f_0} r^{*2} + \frac{f_0 f_4 - 3f_2^2}{4! f_0^2} r^{*4} + \order{r^{*6}}.
\end{align}
To ensure the continuities of $\varphi$ and its derivatives at the center, it is concluded that the choice $q=0$, i.e. the Boulware vacuum, is preferred. In this case, there is no singular value in the RSET~\eqref{eq:RSET_SSS_2d} unlike the Schwarzschild case.

This is the consequence of the conformally flat feature~\eqref{eq:SSS_2d} in the two-dimensional spacetime. In the case of this horizonless regular spacetime, the global structure is conformally flat. Thus we can extend various vacuum states in Minkowski spacetime into this horizonless spacetime by just adding the inhomogeneous geometry term $\log f$ appearing in Eq.~\eqref{eq:phi_2d_sss}, which corresponds to the conformal transformation into the Minkowski spacetime. The Boulware vacuum we have obtained here can be understood as the extension of the Minkowski vacuum which gives the nonsingular $\varphi$ and RSET in the flat spacetime. This is in contrast to the Schwarzschild case where the conformal flatness is limited only to the exterior of the horizon, and the existence of the horizon (or nontrivial geometry inside the horizon) is reflected as the singular behaviors of $\varphi$ and RSET.

\section{Four-dimensional case \label{sec:4D}}
\subsection{The local action for trace anomaly \label{sec:4D_general}}

We commence this section with a review of the local field formulation \cite{Mottola:2006ew} in general four-dimensional spacetime. In four-dimensional spacetime, the trace anomaly for massless particles is known to be \cite{Capper:1974ic, Capper:1975ig, Deser:1976yx, Duff:1977ay, Birrell:1982ix}
\footnote{\label{footnote:anomaly-ambiguity}
In general, the $\Box R$ term or a contribution from a gauge field may appear in the trace anomaly. The $\Box R$ term can be absorbed by the redefinition of a local geometrical counterterm in the effective action, i.e., $R^2$ term, and such a redefinition is equivalent to a field redefinition of the metric, which does not change the later discussion of boundary condition at the origin. A contribution from a gauge field to the trace anomaly, e.g., the invariant of the field strength $\mathrm{tr}\qty[F_{ab}F^{ab}]$ appears if a massless field interacts with the gauge field. These terms will be disregarded in the subsequent discussion for simplicity.
}
\begin{align}
    \ev{ T^\alpha_\alpha }
    = b F+b^{\prime}\left(E-\frac{2}{3} \square R\right),
    \label{eq:Anom4d}
\end{align}
where $E, F$ are the scalar invariant composed of curvature tensors
\begin{align}
    &E \equiv
    { }^* R_{\mu\nu\lambda\sigma}
    { }^* R^{\mu\nu\lambda\sigma}=R_{\mu\nu\lambda\sigma} R^{\mu\nu\lambda\sigma}-4 R_{\mu\nu} R^{\mu\nu}+R^2, \\
    &F \equiv C_{\mu\nu\lambda\sigma} C^{\mu\nu\lambda\sigma}
    =R_{\mu\nu\lambda\sigma} R^{\mu\nu\lambda\sigma}-2 R_{\mu\nu} R^{\mu\nu}+\frac{R^2}{3},
\end{align}
and coefficients $b, b'$ are given as
\begin{align}
    &b=\frac{\hbar}{120(4 \pi)^2}\left(N_S+6 N_F+12 N_V\right), \nonumber \\
    &b^{\prime}=-\frac{\hbar}{360(4 \pi)^2}\left(N_S+11 N_F+62 N_V\right).
    \label{eq:coeff_b}
\end{align}
with the numbers $N_S, N_F,$ and $N_V$ of the massless fields of spin-$0, 1/2,$ and $1$ respectively. 

As in the two-dimensional case, the contribution of such an anomaly in four-dimensional spacetime is taken into account by introducing the following local anomalous action \cite{Shapiro:1994ww, Balbinot:1999vg, Balbinot:1999ri, Mottola:2006ew}
\begin{align}
    S^{(4)}_{\mathrm{anom}}[g ; \varphi, \psi]
    = b' S^{(E)}_{\mathrm{anom}}[g ; \varphi, \psi]
    + b S^{(F)}_{\mathrm{anom}}[g ; \varphi, \psi]
    \label{eq:Laction4d}
\end{align}
with
\begin{align}
    S^{(E)}_{\mathrm{anom}}[g ; \varphi, \psi] 
    & \equiv \frac{1}{2} \int \dd[4]{x} \sqrt{-g}
    \qty[
        -\qty(\square \varphi)^2
        +2\qty(R^{\mu\nu}-\frac{R}{3} g^{\mu\nu})
        \qty(\nabla_\mu\varphi)\qty(\nabla_\nu \varphi)
        +\qty(E-\frac{2}{3} \square R) \varphi
    ], \nonumber \\
    S^{(F)}_{\mathrm{anom}}[g ; \varphi, \psi] 
    & \equiv \int \dd[4]{x} \sqrt{-g}
    \qty[
        -\qty(\square \varphi)\qty(\square \psi)
        +2\qty(R^{\mu\nu}-\frac{R}{3} g^{\mu\nu})
        \qty(\nabla_\mu\varphi)\qty(\nabla_\nu \psi)
        +\frac{1}{2} F\varphi 
        +\frac{1}{2}\qty(E-\frac{2}{3} \square R) \psi
    ].
\end{align}
The field equations for the scalar fields $\varphi$ and $\psi$ are
\begin{align}
    &\Delta_4 \varphi=\frac{1}{2}\left(E-\frac{2}{3} \square R\right), 
    \quad 
    \qty(\Delta_4 \equiv \square^2+2 R^{\mu \nu} \nabla_\mu \nabla_\nu-\frac{2}{3} R \, \square+\frac{1}{3}\left(\nabla^\mu R\right) \nabla_\mu)
    \label{eq:EOM_phi4d}
    \\
    &\Delta_4 \psi=\frac{1}{2} F,
    \label{eq:EOM_psi4d}
\end{align}
which are the fourth-order differential equations in the present four-dimensional case. Solving these equations with Green's function of the operator $\Delta_4$ and substituting the solutions into Eq.~\eqref{eq:Laction4d}, the nonlocal form action can be obtained.
The corresponding RSET is \cite{Mazur:2001aa, Mottola:2006ew, Antoniadis:1995dy, Balbinot:1999ri, Balbinot:1999vg}
\begin{align}
    T_{\mu\nu}^{(4)}[g ; \varphi, \psi]
    = b' E_{\mu\nu} + b F_{\mu\nu},
    \label{eq:4DRSET_general}
\end{align}
where 
\begin{align}
    E_{\mu \nu}
    = & -2\left(\nabla_{(\mu} \varphi\right)\left(\nabla_{\nu)} \square \varphi\right)
    +2 \nabla^\alpha\left[\left(\nabla_\alpha \varphi\right)\left(\nabla_\mu \nabla_\nu \varphi\right)\right]
    -\frac{2}{3} \nabla_\mu \nabla_\nu\left[\left(\nabla_\alpha \varphi\right)\left(\nabla^\alpha \varphi\right)\right]
    \nonumber \\
    &
    +\frac{2}{3} R_{\mu \nu}\left(\nabla_\alpha \varphi\right)\left(\nabla^\alpha \varphi\right) 
    -4 R\indices{^{\alpha}_{(\mu}}\left(\nabla_{\nu)} \varphi\right)\left(\nabla_\alpha \varphi\right)
    +\frac{2}{3} R\left(\nabla_\mu \varphi\right)\left(\nabla_\nu \varphi\right)
    -\frac{2}{3} \nabla_\mu \nabla_\nu \square \varphi \nonumber \\
    &+\frac{1}{6} g_{\mu \nu}\left\{-3(\square \varphi)^2+\square\left[\left(\nabla_\alpha \varphi\right)\left(\nabla^\alpha \varphi\right)\right]+2\left(3 R^{\alpha\beta}-R^{\alpha\beta}\right)\left(\nabla_\alpha \varphi\right)\left(\nabla_\beta \varphi\right)\right\} \nonumber \\
    & -4 C\indices{_\mu^\alpha_\nu^\beta}\nabla_\alpha \nabla_\beta \varphi
    -4 R\indices{^\alpha_{(\mu}} \nabla_{\nu)} \nabla_\alpha \varphi
    +\frac{8}{3} R_{\mu \nu} \square \varphi
    +\frac{4}{3} R \nabla_\mu \nabla_\nu \varphi
    -\frac{2}{3}\left(\nabla_{(\mu} R\right) \nabla_{\nu)} \varphi \nonumber \\
    & +\frac{1}{3} g_{\mu \nu}\left\{2 \square^2 \varphi+6 R^{\alpha\beta} \nabla_\alpha \nabla_\beta \varphi-4 R \square \varphi+\left(\nabla^\alpha R\right) \nabla_\alpha \varphi\right\}, \\[8pt]
    F_{\mu \nu}
    = & -2\left(\nabla_{(\mu} \varphi\right)\left(\nabla_{\nu)} \square \psi\right)
    -2\left(\nabla_{(\mu} \psi\right)\left(\nabla_{\nu)} \square \varphi\right)
    +2 \nabla^\alpha\left[\left(\nabla_\alpha \varphi\right)\left(\nabla_\mu \nabla_\nu \psi\right)
    +\left(\nabla_\alpha \psi\right)\left(\nabla_\mu \nabla_\nu \varphi\right)\right] \nonumber \\
    & 
    -\frac{4}{3} \nabla_\mu \nabla_\nu\left[\left(\nabla_\alpha \varphi\right)\left(\nabla^\alpha \psi\right)\right]
    +\frac{4}{3} R_{\mu \nu}\left(\nabla_\alpha \varphi\right)\left(\nabla^\alpha \psi\right)
    -4 R\indices{^\alpha_{(\mu}}\left[\left(\nabla_{\nu)} \varphi\right)\left(\nabla_\alpha \psi\right)+\left(\nabla_{\nu)} \psi\right)\left(\nabla_\alpha \varphi\right)\right] \nonumber \\
    & +\frac{4}{3} R\left(\nabla_{(\mu} \varphi\right)\left(\nabla_{\nu)} \psi\right) 
    -4 \nabla_\alpha \nabla_\beta \left(C\indices{_{(\mu}^\alpha_{\nu)}^\beta}\varphi\right)  
    -2 C\indices{_\mu^\alpha_\nu^\beta} R_{\alpha\beta} \varphi
    -\frac{2}{3} \nabla_\mu \nabla_\nu \square \psi 
    -4 C\indices{_\mu^\alpha_\nu^\beta} \nabla_\alpha \nabla_\beta \psi
    \nonumber \\
    & +\frac{1}{3} g_{\mu \nu}\left\{-3(\square \varphi)(\square \psi)
    +\square\left[\left(\nabla_\alpha \varphi\right)\left(\nabla^\alpha \psi\right)\right]+2\left(3 R^{\mu\nu}-R^{\mu\nu}\right)\left(\nabla_\alpha \varphi\right)\left(\nabla_\nu \psi\right)\right\}
    \nonumber \\
    & -4 R_{(\mu}^\alpha\left(\nabla_{\nu)} \nabla_\alpha \psi\right)
    +\frac{8}{3} R_{\mu \nu} \square \psi
    +\frac{4}{3} R \nabla_\mu \nabla_\nu \psi
    -\frac{2}{3}\left(\nabla_{(\mu} R\right) \nabla_{\nu)} \psi
    \nonumber \\
    & 
    +\frac{1}{3} g_{\mu \nu}\left\{2 \square^2 \psi+6 R^{\alpha\beta} \nabla_\alpha \nabla_\beta \psi
    -4 R \square \psi+\left(\nabla^\alpha R\right)\left(\nabla_\alpha \psi\right)\right\} .
\end{align}
One can check that the trace of this RSET properly reproduces Eq.~\eqref{eq:Anom4d}. In principle, there is an ambiguity in the definition of the anomaly-induced effective action due to conformally invariant functionals. Additionally, local higher curvature corrections may appear in the total gravitational effective action after renormalization (see footnote~\ref{footnote:anomaly-ambiguity}). For this moment, we do not consider these contributions.

The four-dimensional anomalous action~\eqref{eq:Laction4d} is invariant under the constant shift $\psi \rightarrow \psi + \psi_0$.
This fact implies the existence of the Noether current
\begin{align}
    J^\mu
    \equiv \nabla^\mu \square \varphi+2\left(R^{\mu\nu}-\frac{R}{3} g^{\mu\nu}\right) \nabla_\nu \varphi-\frac{1}{2} \Omega^\mu+\frac{1}{3} \nabla^\mu R,
    \label{eq:current_4D}
\end{align}
with topological current whose divergence is the four-dimensional Euler-Gauss-Bonnet integrand
\begin{align}
    \nabla_\alpha \Omega^\alpha=E=R_{\alpha\beta\gamma\epsilon} R^{\alpha\beta\gamma\epsilon}-4 R_{\alpha\beta} R^{\alpha\beta}+R^2.
\end{align}
Also, the corresponding charge 
\begin{align}
    \int_{\Sigma} J^a \dd{\Sigma}_a,
\end{align}
on the Cauchy surface $\Sigma$ is conserved during time evolution. This charge parametrizes the homogenous solutions of $\varphi$ as in the two-dimensional case. 
It should be noted that the integration on timelike or null boundaries must be taken into account if it is relevant. Specifically, for a static asymptotically flat spacetime, the following definition is useful:
\begin{align}
    \tilde{q}= \frac{1}{4\pi} \int_{\partial\Sigma} J^a \dd{\Sigma}_a.
    \label{eq:charge_4D}
\end{align}
where $\partial\Sigma$ is two-dimensional boundaries of $\Sigma$. To accord with the notation in \cite{Mottola:2006ew}, we will use the dimensionless version $q = 2M\tilde{q}$ of this definition in later examples of the Schwarzschild and a horizonless Bardeen-type spacetimes.

On the other hand, the action~\eqref{eq:Laction4d} is not invariant in general under the constant shift $\varphi \rightarrow \varphi + \varphi_0$ corresponding to the global Weyl rescaling of the metric, unless 
\begin{align*}
    F = 0.
\end{align*}
It is shown that the anomalous action~\eqref{eq:Laction4d} scales linearly (or logarithmically in length scale) under the global Weyl transformation as well as the constant shift of $\varphi$. Thus it is relevant even at large distances.

\subsection{Stress-energy tensor in the Schwarzschild spacetime \label{sec:4D_Sch}}
Because of the conformally flat structure of the near horizon geometry in the four-dimensional Schwarzschild spacetime, the anomalous action~\eqref{eq:Laction4d} and the field equations~\eqref{eq:EOM_phi4d} and \eqref{eq:EOM_psi4d} provide an accurate description of the system. Since the Schwarzschild spacetime is Ricci-flat and thus these two field equations coincide, the (derivative of) static solution $\varphi$ and $\psi$ is found to have the same analytical form  \cite{Balbinot:1999vg, Balbinot:1999ri, Mottola:2006ew}
\begin{align}
    &\dv{\varphi}{r}
    = \frac{q-2}{6 M}
    \left(\frac{r}{2 M}+1+\frac{2 M}{r}\right) 
    \log \left(1-\frac{2 M}{r}\right) 
    -\frac{q}{6 r}\left[\frac{4 M}{r-2 M} \log \left(\frac{r}{2 M}\right)
    +\frac{r}{2 M}+3\right] \nonumber \\
    & \hspace{200pt}
    -\frac{1}{3 M}-\frac{1}{r} 
    +\frac{2 M c_{\rm H}}{r(r-2 M)}
    +\frac{c_{\infty}}{2 M}\left(\frac{r}{2 M}+1+\frac{2 M}{r}\right), 
    \label{eq:phi_4dSch} \\[5pt]
    &\dv{\psi}{r}
    = \frac{q'-2}{6 M}
    \left(\frac{r}{2 M}+1+\frac{2 M}{r}\right) 
    \log \left(1-\frac{2 M}{r}\right) 
    -\frac{q'}{6 r}\left[\frac{4 M}{r-2 M} \log \left(\frac{r}{2 M}\right)
    +\frac{r}{2 M}+3\right] \nonumber \\
    & \hspace{200pt}
    -\frac{1}{3 M}-\frac{1}{r} 
    +\frac{2 M d_{\rm H}}{r(r-2 M)}
    +\frac{d_{\infty}}{2 M}\left(\frac{r}{2 M}+1+\frac{2 M}{r}\right), 
    \label{eq:psi_4dSch}
\end{align}
with six constants $\qty{q, c_{\rm H}, c_\infty;~q', d_{\rm H}, d_\infty}$ to parametrize the homogeneous parts. One can check that this $q$ is identified with the one that appeared as a Noether charge~\eqref{eq:charge_4D}, which implies a topological defect at the origin as in the Minkowski case. The remaining $c_{\rm H}, d_{\rm H}$ control the leading behaviors at $r=2M$ and $c_\infty, d_\infty$ for $r\rightarrow\infty$ respectively. In addition to these spatial terms, a linear term of $t$ can be added to the solution  
\begin{align}
    &\varphi(r,t) = \varphi(r) + \frac{p}{2M} t,
    \label{eq:phi_4dSch_t} \\
    &\psi(r,t) = \psi(r) + \frac{p'}{2M} t.
    \label{eq:psi_4dSch_t}
\end{align}

The eight constants $\{p, q, c_{\rm H}, c_\infty;~p', q', d_{\rm H}, d_\infty \}$ determine the homogeneous, i.e. vacuum-dependent, parts of $\varphi(r, t)$ and $\psi(r, t)$. The Boulware vacuum \cite{Boulware:1974dm, Boulware:1975fe}, which is static and reduces to the Minkowski vacuum for $r\rightarrow\infty$, is given by
\begin{align}
    p=q=c_\infty=p'=q'=d_\infty=0,
\end{align}
so that the terms in Eqs.~\eqref{eq:phi_4dSch} and \eqref{eq:phi_4dSch_t}, as well as the resulting RSET, which diverge for the far limit, vanish. The remaining $c_{\rm H}, d_{\rm H}$ cannot eliminate all divergent behaviors for the $r \rightarrow 2M$ limit. Therefore, the auxiliary fields and the resulting RSET blow up in this limit as shown in \cite{Brown:1986jy, Brown:1986tj, Mottola:2006ew}.

\subsection{Proper vacuum state and RSET in static horizonless spacetime with spherical symmetry \label{sec:4D_horizonless}}

We consider the case of a regular and asymptotically flat spacetime without any horizons. The local field formulation provides a clear method for analyzing the RSET in general spacetimes. To find the vacuum state and corresponding RSET, all we have to do is to solve the field equations \eqref{eq:EOM_phi4d} and \eqref{eq:EOM_psi4d} and calculate the RSET~\eqref{eq:4DRSET_general}. By taking advantage of this method, we will demonstrate, through a case study, how the regularities of the solutions can indicate the proper vacuum state for the four-dimensional horizonless spacetime. 

Well-known examples of regular metrics include the Hayward-type \cite{Hayward:2005gi, Carballo-Rubio:2022nuj} and Bardeen-type \cite{1968qtr..conf...87B, Carballo-Rubio:2022nuj} geometries. Regarding the Hayward-type geometry, the higher derivatives of the metric have discontinuities at the center, as in \appref{sec:hayward}. This, in turn, causes the higher derivatives of the scalar fields $\varphi$ and $\psi$ to become irregular through the field equations. Therefore, we will investigate the Bardeen-type geometry case in the following.
\footnote{
As a horizonless spacetime, one can consider the scenario where the background classical matter distribution has discontinuities, such as a thin shell. In this case, we need to solve the junction conditions for the higher derivatives of the fields, as provided in \appref{sec:JC}. 
Addressing the junction conditions in the shell model is more challenging and complex than in smooth spacetime; therefore, we will not pursue it further.
}




\subsubsection{Case-study: Bardeen-type}

In the following, we demonstrate how the regularity of the auxiliary fields $\varphi$ and $\psi$, as well as the resulting RSET, lead to the preferred vacuum state using the Bardeen-type horizonless spacetime \cite{1968qtr..conf...87B, Carballo-Rubio:2022nuj} as an illustrative example. The Bardeen-type metric was originally proposed to describe a regular black hole without the central singularity \cite{1968qtr..conf...87B}, and its metric is given as
\begin{align}
    \dd{s^2}
    =-f(r) \dd{t}^2+  f(r)^{-1} \dd{r}^2
    +r^2 \dd{\Omega}^2,
    \label{eq:Bardeen_met}
\end{align}
where
\begin{align}
    f(r)=1-\frac{2 m(r)}{r}, \quad m(r)=\frac{M r^3}{\qty(r^2+R^2)^{3/2}},
\end{align}
with arbitrary parameters $M$ and $R$.
For sufficiently distant regions, $r \gg M \text{~and~} R$, the geometry reduces to the Schwarzschild spacetime with the ADM mass $M$. Whether the spacetime has the event horizon is determined by a critical mass $M_\mathrm{crit} / R = 3\sqrt{3}/4$, and is classified as follows,
\begin{align}
    \begin{cases}
        M > M_\mathrm{crit}
        \quad \text{a regular BH with two horizons} \\[5pt]
        M = M_\mathrm{crit}
        \quad \text{a regular BH with one degenerate horizon} \\[5pt]
        M < M_\mathrm{crit}
        \quad \text{a horizonless regular spacetime} 
    \end{cases}
\end{align}
Since we are now interested in the semiclassical effect on the horizonless regular spacetime, we focus on the case with $M < M_\mathrm{crit}$. Around the center $r=0$ the curvature reduces to
\footnote{This parity-even nature of the spacetime ensures that the fields $\varphi, \psi$ as in Eq.~\eqref{eq:scsol_center} should not exhibit any non-smooth behaviors in their derivatives. See also \appref{sec:hayward} for further discussion on this point.}
\begin{align}
    \mathcal{R}(r) = \frac{24 M}{R^3}-\frac{90 M r^2}{R^5}+\frac{210 M r^4}{R^7}-\frac{1575 M r^6}{4 R^9} + \order{r^7}.
\end{align}
The geometry is regular and parity-even around $r=0$, and thus the resulting scalar fields $\varphi$ and $\psi$ should also be. 

\begin{figure}[tp]%
  \begin{minipage}[t]{0.5\linewidth}%
    \centering%
    \includegraphics[keepaspectratio, width=0.97\linewidth]{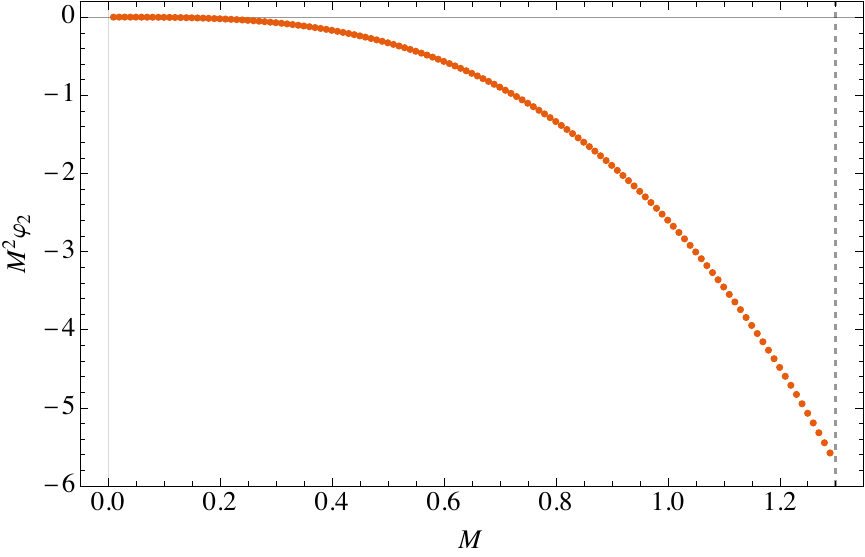}%
    \subcaption{$\varphi_2$--$M$}%
    \label{fig:bd_phi_2}%
  \end{minipage}%
  \begin{minipage}[t]{0.5\linewidth}%
    \centering%
    \includegraphics[keepaspectratio, width=\linewidth]{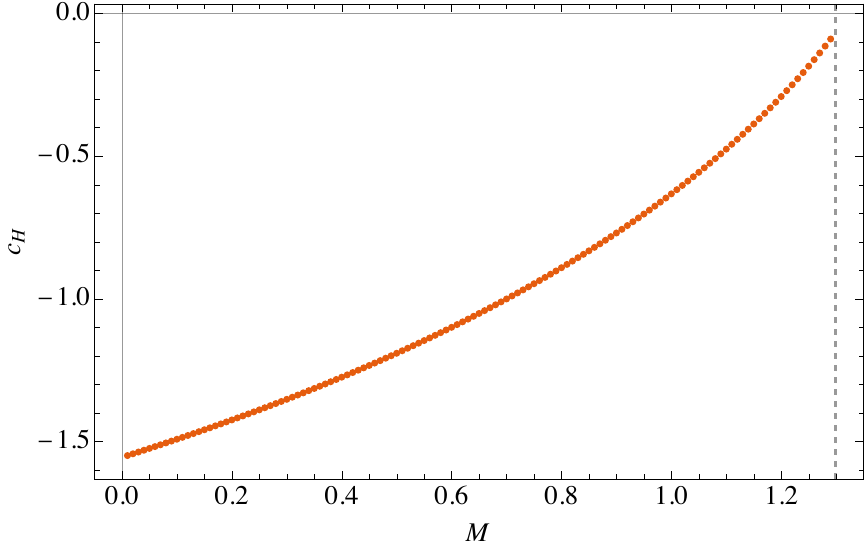}%
    \subcaption{$c_H$--$M$}%
    \label{fig:bd_cH_inf}%
  \end{minipage}%
  \\[10pt]
  \begin{minipage}[t]{0.48\linewidth}%
    \centering%
    \includegraphics[keepaspectratio, width=\linewidth]{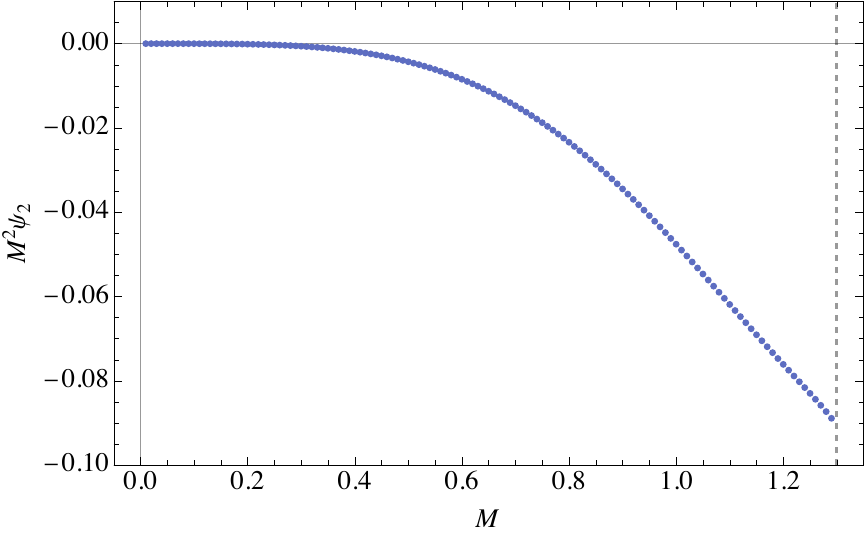}%
    \subcaption{$\psi_2$--$M$}%
    \label{fig:bd_psi_2}%
  \end{minipage}%
  \hspace{5pt}
  \begin{minipage}[t]{0.48\linewidth}%
    \centering%
    \includegraphics[keepaspectratio, width=\linewidth]{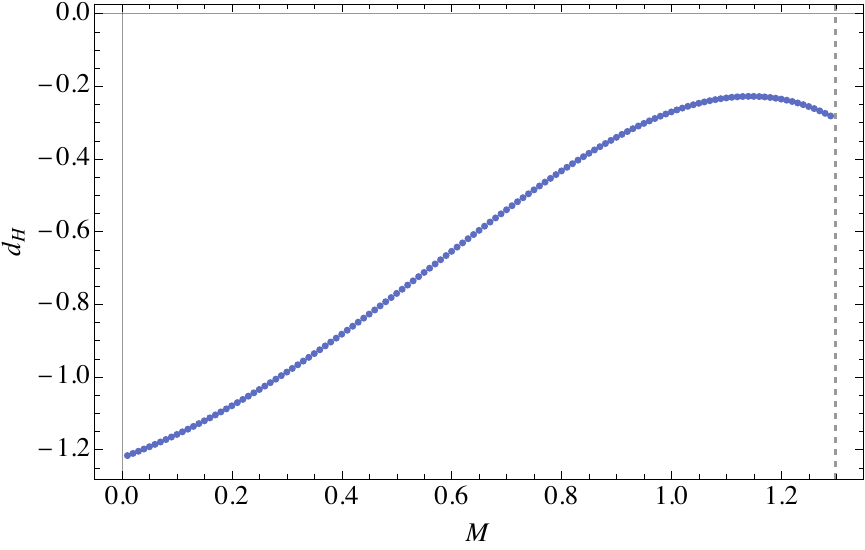}%
    \subcaption{$d_H$--$M$}%
    \label{fig:bd_dH_inf}%
  \end{minipage}%
  \\[10pt]
  \begin{minipage}[t]{0.5\linewidth}%
    \centering%
    \includegraphics[keepaspectratio, width=\linewidth]{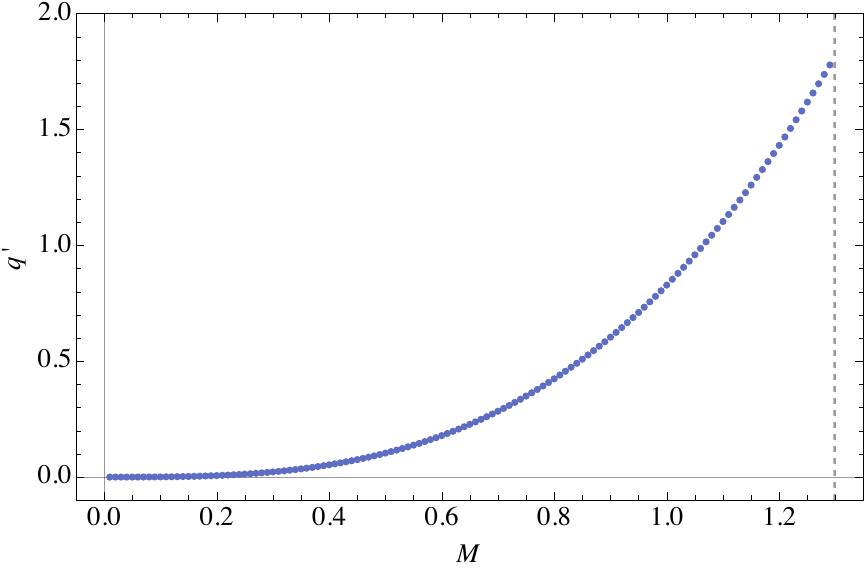}%
    \subcaption{$q'$--$M$}%
    \label{fig:bd_q'_inf}%
  \end{minipage}%
  \caption{The parameters of the numerical solutions of $\varphi'$ and $\psi'$ in the Bardeen spacetime (normalized by $R$). The vertical dashed line denotes the black hole limit $M=M_{\mathrm{crit}}$. One can see that $q'$ takes nonzero values for larger $M$. This result indicates the regularity condition at the center does not necessarily result in the Boulware vacuum where $q=q'=0$.}
  \label{fig:bd_param_sol}
\end{figure}%

\begin{figure}[tp]%
  \begin{minipage}[t]{0.5\linewidth}%
    \centering%
    \includegraphics[keepaspectratio, width=0.97\linewidth]{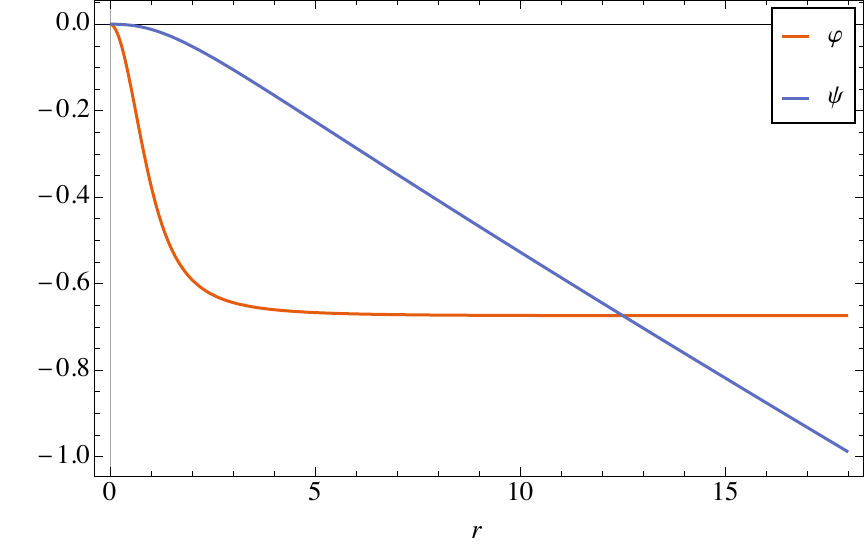}%
    \subcaption{$\varphi(r), \psi(r)$}%
    \label{fig:bd_sc}%
  \end{minipage}%
  \begin{minipage}[t]{0.5\linewidth}%
    \centering%
    \includegraphics[keepaspectratio, width=\linewidth]{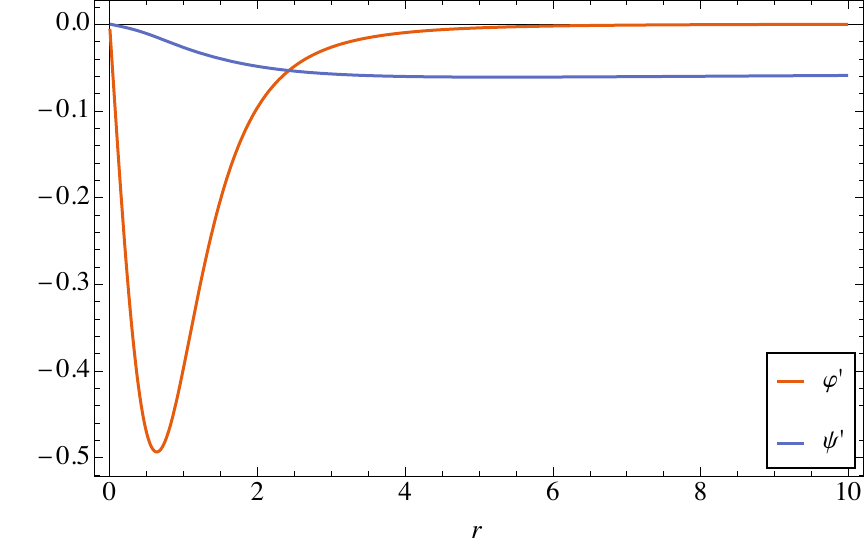}%
    \subcaption{$\varphi'(r), \psi'(r)$}%
    \label{fig:bd_dsc}%
  \end{minipage}%
  \\[10pt]
  \begin{minipage}[t]{0.5\linewidth}%
    \centering%
    \includegraphics[keepaspectratio, width=\linewidth]{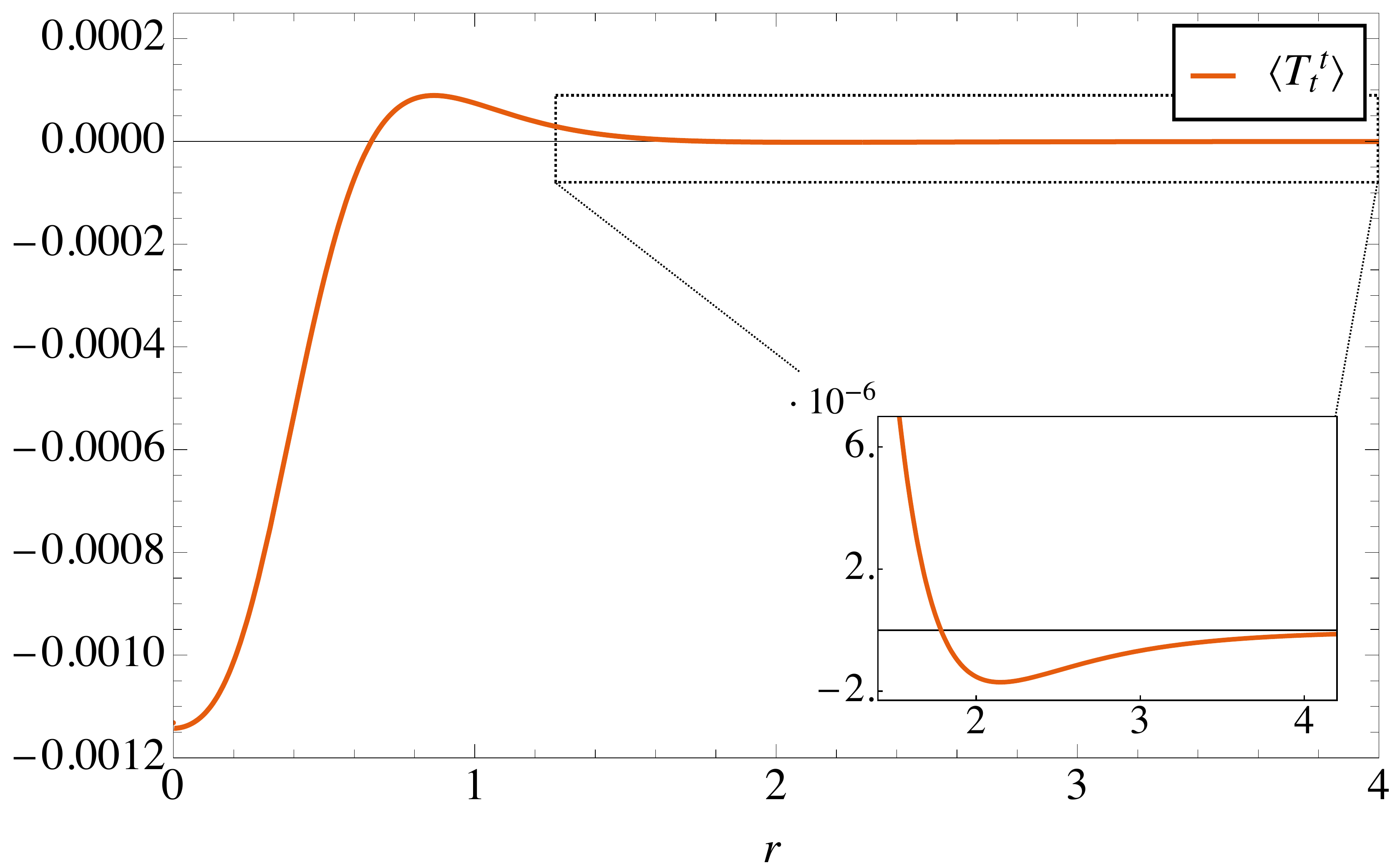}%
    \subcaption{$\ev{T\indices{_t^t}}$}%
    \label{fig:bd_SET00}%
  \end{minipage}%
  \begin{minipage}[t]{0.5\linewidth}%
    \centering%
    \includegraphics[keepaspectratio, width=\linewidth]{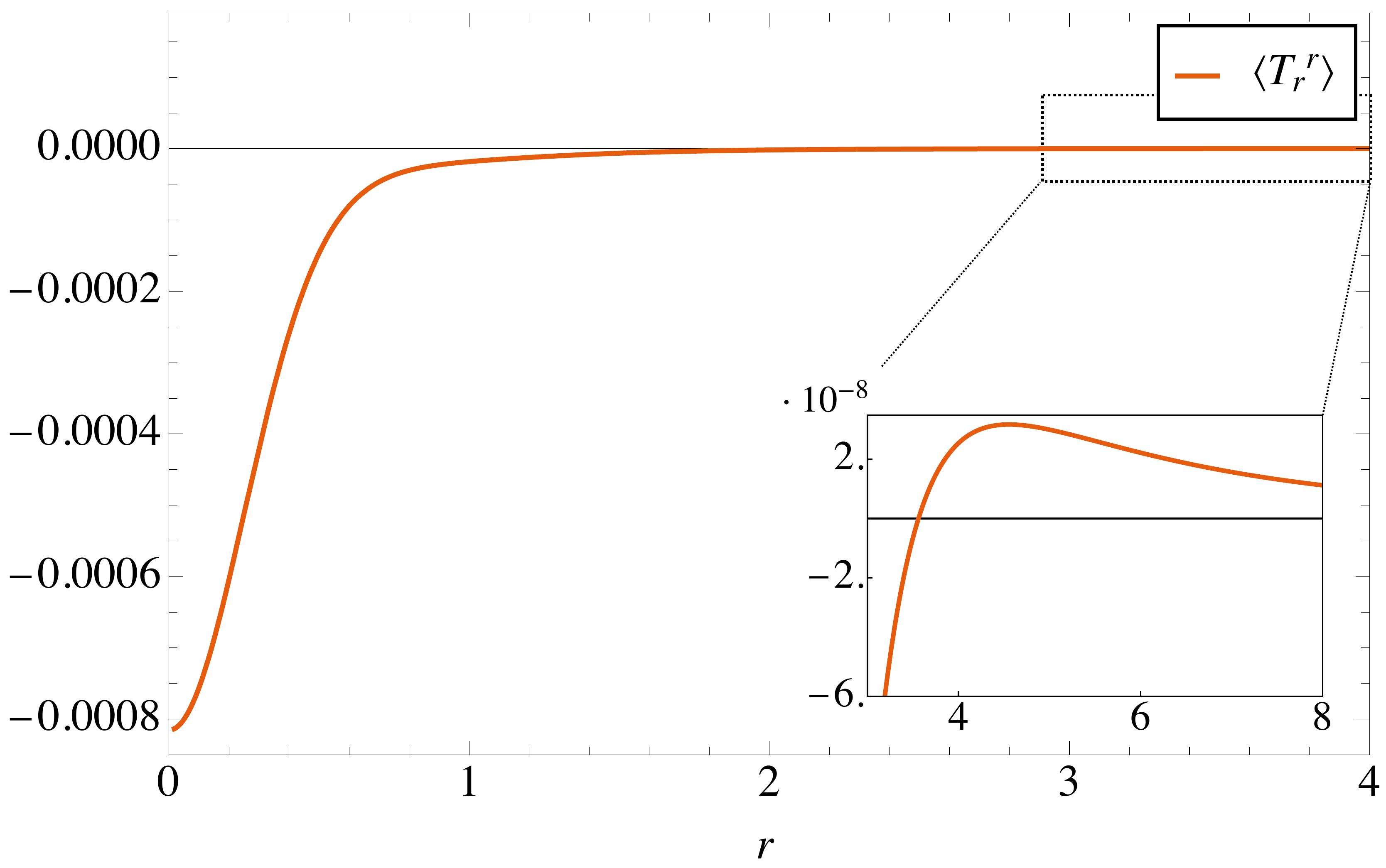}%
    \subcaption{$\ev{T\indices{_r^r}}$}%
    \label{fig:bd_SET11}%
  \end{minipage}%
  \\[10pt]
  \begin{minipage}[t]{0.5\linewidth}%
    \centering%
    \includegraphics[keepaspectratio, width=\linewidth]{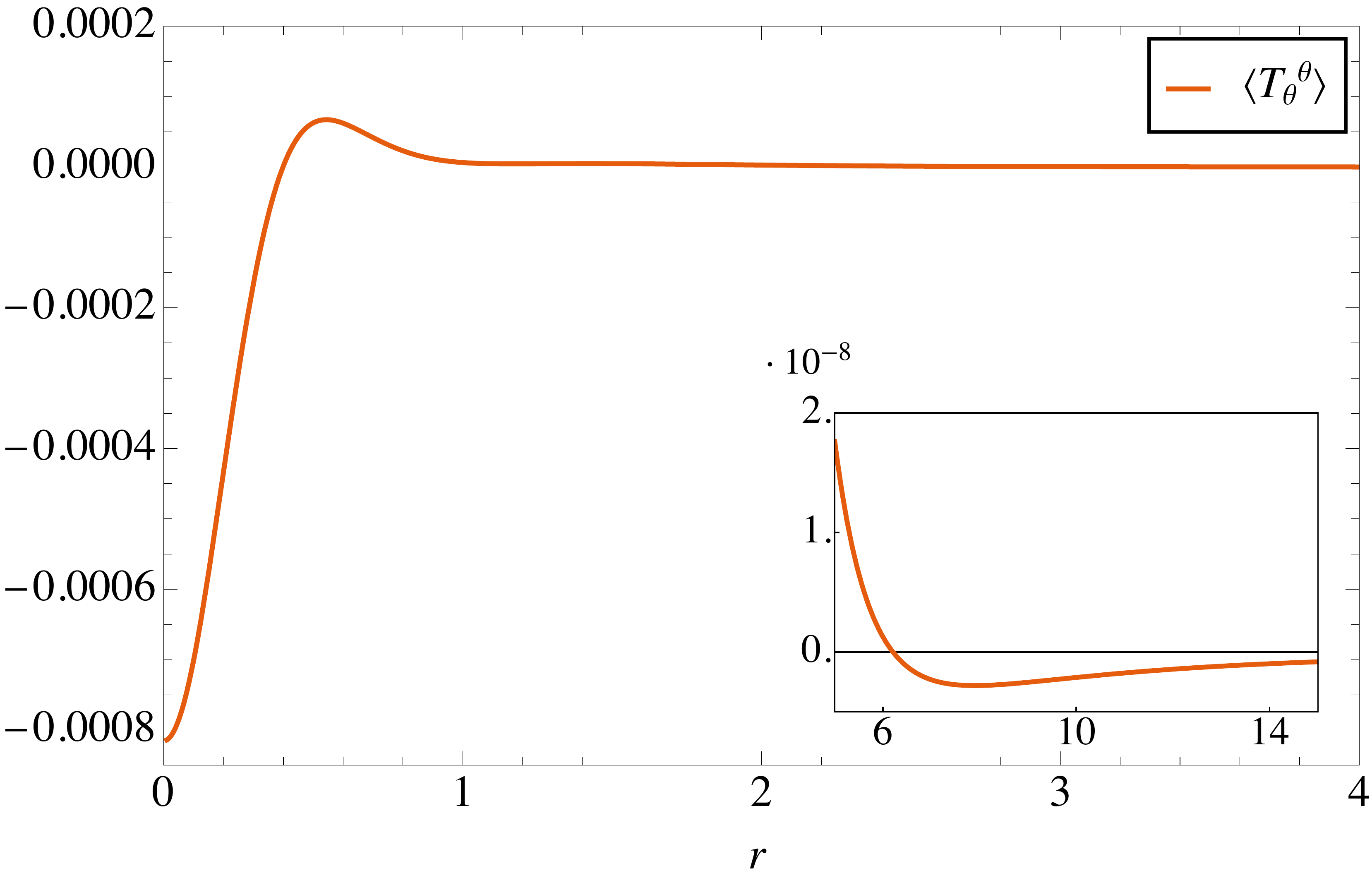}%
    \subcaption{$\ev{T\indices{_\theta^\theta}}$}%
    \label{fig:bd_SET22}%
  \end{minipage}%
  \begin{minipage}[t]{0.5\linewidth}%
    \centering%
    \includegraphics[keepaspectratio, width=\linewidth]{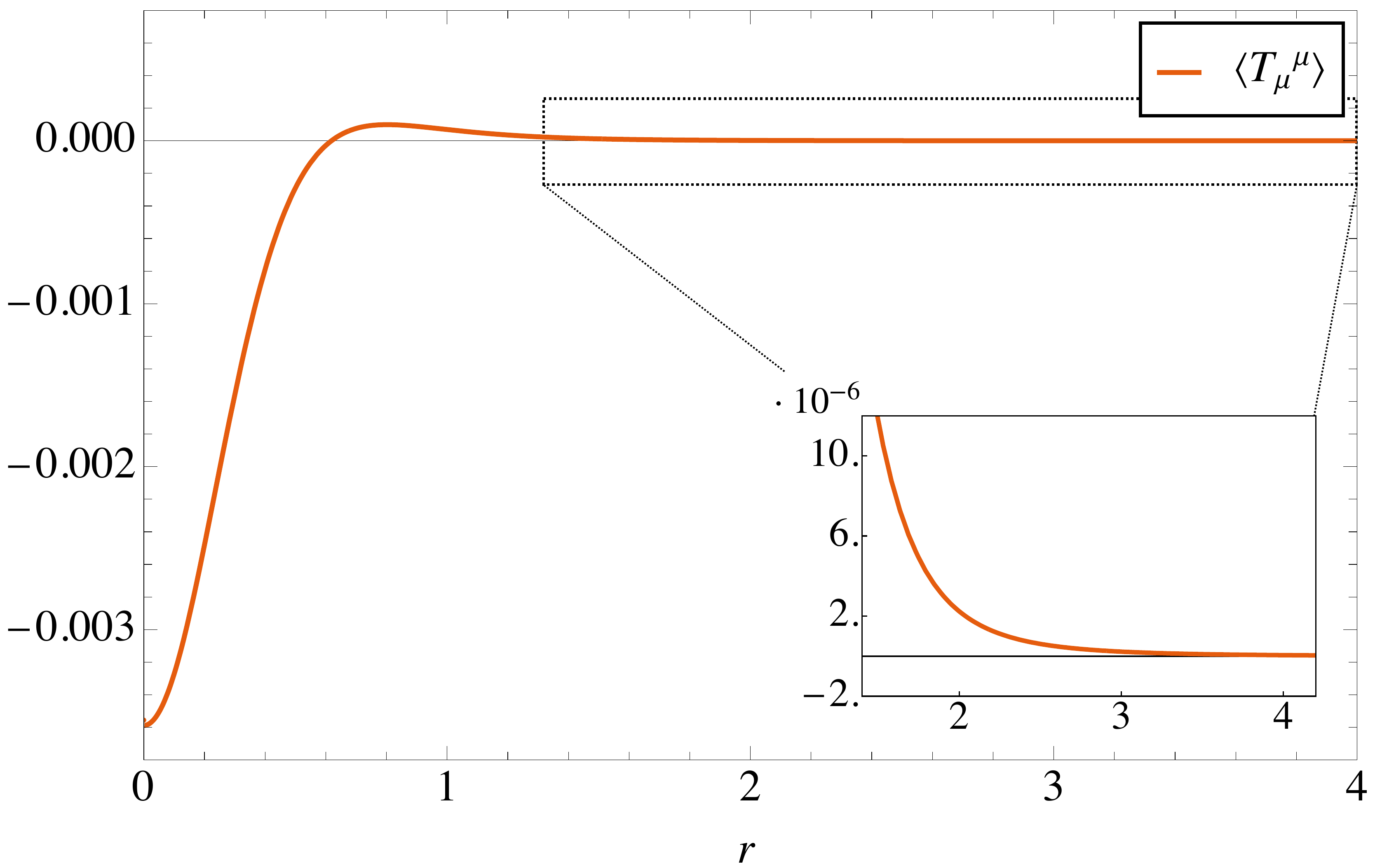}%
    \subcaption{$\ev{T\indices{_\mu^\mu}}$}%
    \label{fig:bd_SETtr}%
  \end{minipage}%
  \caption{ (a), (b) The profiles of the scalar fields $\varphi, \psi$ and their derivatives for $M=0.5 R$. $\psi(r)$ shows linear growth asymptotically, indicating that the vacuum state is no longer Boulware. (c)--(f) Radial dependence of each component and the trace of the RSET. Mini panels zoom in on the asymptotic region. They all indicate nontrivial but regular distribution around the center, and decrease for a distant region.}
  \label{fig:bd_sol}
\end{figure}%

By numerically solving the fourth-order differential equations of the fields~\eqref{eq:EOM_phi4d} and \eqref{eq:EOM_psi4d}, one can obtain the distribution of the auxiliary scalar fields $\varphi$ and $\psi$. Before solving these equations directly, it is helpful to consider the Noether current $J^\mu$ in Eq.~\eqref{eq:current_4D}, which comes from the global Weyl invariance.\footnote{Here we assume the static configuration, so that the scalar fields and the Noether current depend only on the radial coordinate $r$.} 
Since the current is divergence-free
\begin{align}
    \nabla_{\mu} J^{\mu} = 0 \quad
    \Rightarrow \quad \dv{r} J^{r} + \frac{2}{r} J^{r} =0,
\end{align}
the solution satisfying Eq.~\eqref{eq:charge_4D} is derived:
\begin{align}
    J^{r} (r) = \frac{q}{2Mr^2}.
    \label{eq:J_BD}
\end{align}
To avoid the singular behavior at the center, the Noether charge $q$ and thus $J^r$ must vanish. With the definition of the current~\eqref{eq:current_4D}, the reduced third-order differential equation for $\varphi$ is obtained as follows:
\begin{align}
    \nabla^\mu \square \varphi+2\left(R^{\mu\nu}-\frac{R}{3} g^{\mu\nu}\right) \nabla_\nu \varphi-\frac{1}{2} \Omega^\mu+\frac{1}{3} \nabla^\mu R = 0.
    \label{eq:red_EOM_phi4d}
\end{align}
In addition to the differential equation \eqref{eq:EOM_psi4d} for $\psi$, we use the above equation for $\varphi$ instead of Eq.~\eqref{eq:EOM_phi4d} to reduce computational costs. 
These field equations \eqref{eq:red_EOM_phi4d} and \eqref{eq:EOM_psi4d} can be solved numerically as the two-boundary value problem with boundaries at the center $r = 0$ and a distant end point $r=r_{\rm e} \gg \max(2M,R)$. One can see that the auxiliary fields $\varphi$ and $\psi$ without derivatives are absent in the differential equations; hence, they can be viewed as the second-order and the third-order differential equations for the first derivative $\varphi'$ and $\psi'$, respectively. Thus, two and three boundary conditions must be specified to obtain solutions for the auxiliary fields.

Around the center, we assume regular series-type solutions for $\varphi$ and $\psi$:
\begin{align}
    \varphi(r) &= \varphi_0 + \varphi_1 r + \frac{1}{2!} \varphi_2 r^2
    + \frac{1}{3!} \varphi_3 r^3 + \order{r^4}, \nonumber \\
    \psi(r) &= \psi_0 + \psi_1 r + \frac{1}{2!} \psi_2 r^2
    + \frac{1}{3!} \psi_3 r^3 + \order{r^4}.
    \label{eq:scsol_center}
\end{align}
To avoid discontinuities in their derivatives, we must demand the following (2+2) conditions
\begin{align}
    \varphi_1 = \varphi_3 = \psi_1 = \psi_3 = 0.
    \label{eq:bd_BC1}
\end{align}
The second derivatives at the center, $\varphi_2$ and $\psi_2$, are used as shooting parameters to be determined from the consistency. 

For sufficiently distant regions, the metric reduces to the Schwarzschild one for the distant region $r \simeq r_{\rm e}$; and hence, Eqs.~\eqref{eq:phi_4dSch} and \eqref{eq:psi_4dSch} can be used as the asymptotic solutions. In this region, Eqs.~\eqref{eq:phi_4dSch} and \eqref{eq:psi_4dSch} are expanded as
\begin{align}
    &\dv{\varphi}{r}
    = \frac{c_{\infty} r}{4 M^2} + \frac{2c_{\infty} - q}{4 M} + \frac{c_{\infty} - q}{r} -\frac{2M}{3 r^2} q \log \left(\frac{r}{2M}\right) + \frac{2M}{r^2} \left[ c_{\rm H} -\frac{11}{18}(q-2) \right] + \cdots, \quad 
    (r \rightarrow \infty)
    ,
    \label{eq:phi_4dSch_expansion} \\[5pt]
    &\dv{\psi}{r}
    = \frac{d_{\infty} r}{4 M^2} + \frac{2d_{\infty} - q'}{4 M} + \frac{d_{\infty} - q'}{r} -\frac{2M}{3 r^2} q' \log \left(\frac{r}{2M}\right) + \frac{2M}{r^2} \left[ d_{\rm H} -\frac{11}{18}(q'-2) \right] + \cdots, \quad 
    (r \rightarrow \infty).
    \label{eq:psi_4dSch_expansion}
\end{align}
As is seen in the above equations, the first derivatives of the auxiliary fields generically blow up for $r\rightarrow\infty$. To eliminate such browing-up behavior, we demand the following condition,
\begin{align}
    c_\infty = d_\infty = 0.
    \label{eq:bd_BC2}
\end{align}
To obtain a static configuration like the Boulware state, we need to set 
\begin{align}
    p = p' = 0.
    \label{eq:bd_pp}
\end{align}
As is seen before, we also have the condition from the finiteness for the Noether current~\eqref{eq:J_BD},
\begin{align}
    q=0.
    \label{eq:bd_BC3}
\end{align} 
The three constants $c_{\rm H},~d_{\rm H}$, and $q'$ are to be determined.

It should be noted that the boundary conditions~\eqref{eq:bd_BC1} and \eqref{eq:bd_BC2} contain redundant ones; the form of the Noether current~\eqref{eq:J_BD} already prevents nonzero $\varphi_1, \varphi_3,$ and $c_\infty$, as is confirmed by substituting the asymptotic solution~\eqref{eq:scsol_center} or \eqref{eq:phi_4dSch} into it. 

In summary, what we solve are the second-order ordinary differential equation \eqref{eq:red_EOM_phi4d} for $\varphi'$, and the third-order one~\eqref{eq:EOM_psi4d} for $\psi'$. As boundary conditions, we have redundant ones $\varphi_1 = \varphi_3 = c_\infty= 0$, and additional condition $q=0$ for the former field equation. Also, we have $\psi_1 = \psi_3 = d_\infty = 0$ for the latter equation. We solve this system as a two-boundary value problem, and find five shooting parameters, $\varphi_2, c_{\rm H}$ for $\varphi'$, and $\psi_2, d_{\rm H}, q'$ for $\psi'$, which specify the solutions uniquely.  

Under these conditions, we numerically solved Eqs.~\eqref{eq:red_EOM_phi4d} and \eqref{eq:EOM_psi4d} with the shooting method from the two boundaries. By matching two numerical solutions at matching point $r=r_{\rm m}$ with sufficient accuracy, we determined the shooting parameters $\{ \varphi_2, \psi_2, c_{\rm H}, d_{\rm H}, q' \}$ and constructed consistent solutions. The boundaries were placed at $r_{\rm m} = 15 R$ and $r_{\rm e} = 80R$. For numerical implementation, we used the \textit{Mathematica} functions \cite{Mathematica}.

The figures displayed in \figref{fig:bd_param_sol} illustrate the outcomes of the shooting parameters $\varphi_2$, $\psi_2$, $q'$, $c_{\rm H}$, and $d_{\rm H}$ in Eqs.~\eqref{eq:phi_4dSch} and \eqref{eq:scsol_center} for consistent solutions corresponding to each value of $M$. It is noteworthy that $q'$, which controls the linear terms of $\psi(r)$ for large $r$, acquires nonzero values. This specific value of $q'$ indicates the deviation from the Boulware vacuum, which requires $q=q'=0$, but instead leads to a distinct vacuum that ensures regular RSET throughout the spacetime. 
Figure \ref{fig:bd_sol} shows the typical solutions of the scalar fields, their derivatives, and the resulting RSET value for $M=0.5R$. It should be noted that the constant shifts of the scalar fields are undetermined, and thus we have set $\varphi(0)=\psi(0)=0$. Also, we take $N_S=1, N_F = N_V =0$ for the numerical factors~\eqref{eq:coeff_b}. As illustrated in FIGs.~\ref{fig:bd_sc}, \ref{fig:bd_dsc}, $\psi$ exhibit a linear growth in the distant region, indicating that they do not coincide with those in the Minkowski vacuum asymptotically. This profile also means that the Boulware vacuum, where the scalar fields asymptotically vanish, is not a preferred solution in this regular spacetime case. 

\begin{figure}[tp]%
  \begin{minipage}[t]{0.5\linewidth}%
    \centering%
    \includegraphics[keepaspectratio, width=\linewidth]{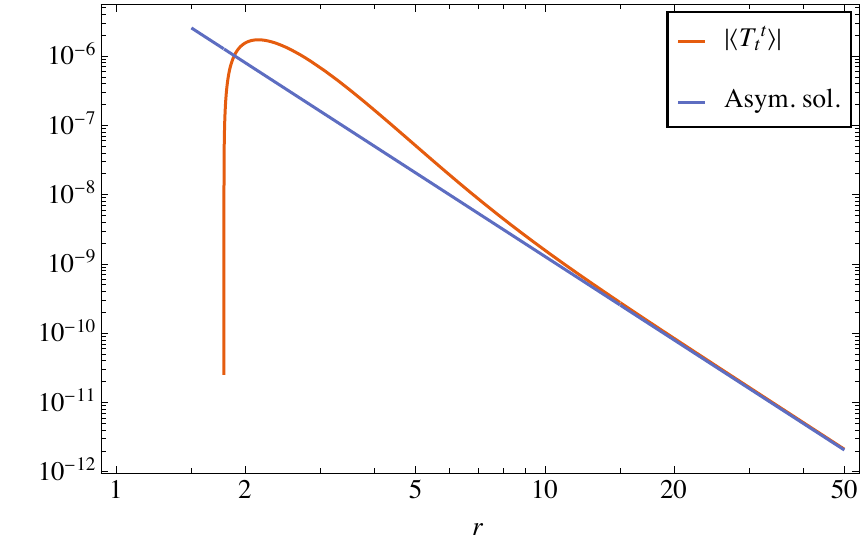}%
    \subcaption{$\ev{T\indices{_t^t}}$}%
    \label{fig:bd_SET00_asym}%
  \end{minipage}%
  \begin{minipage}[t]{0.5\linewidth}%
    \centering%
    \includegraphics[keepaspectratio, width=\linewidth]{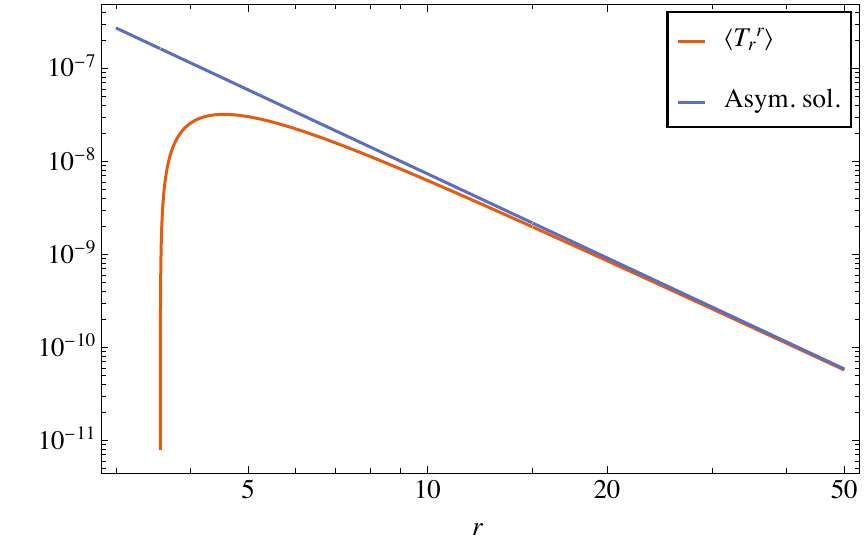}%
    \subcaption{$\ev{T\indices{_r^r}}$}%
    \label{fig:bd_SET11_asym}%
  \end{minipage}%
  \\[10pt]
  \begin{minipage}[t]{0.5\linewidth}%
    \centering%
    \includegraphics[keepaspectratio, width=\linewidth]{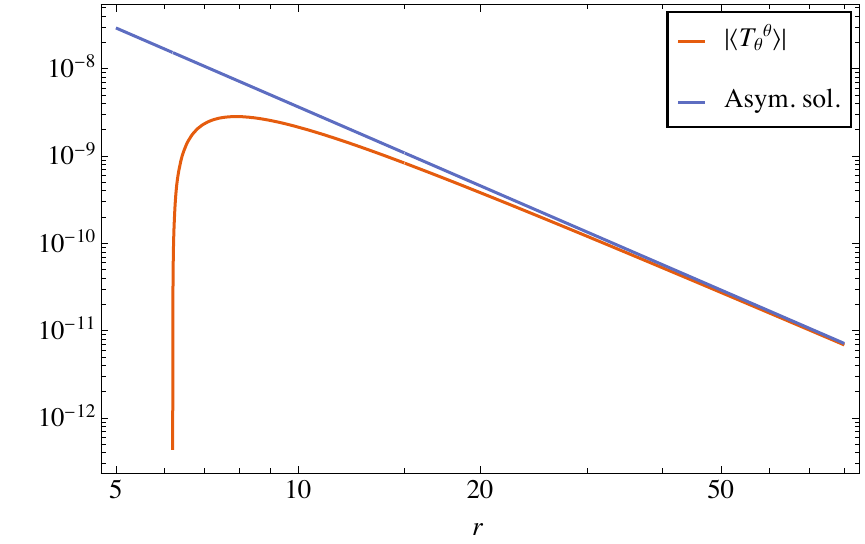}%
    \subcaption{$\ev{T\indices{_\theta^\theta}}$}%
    \label{fig:bd_SET22_asym}%
  \end{minipage}%
  \begin{minipage}[t]{0.5\linewidth}%
    \centering%
    \includegraphics[keepaspectratio, width=\linewidth]{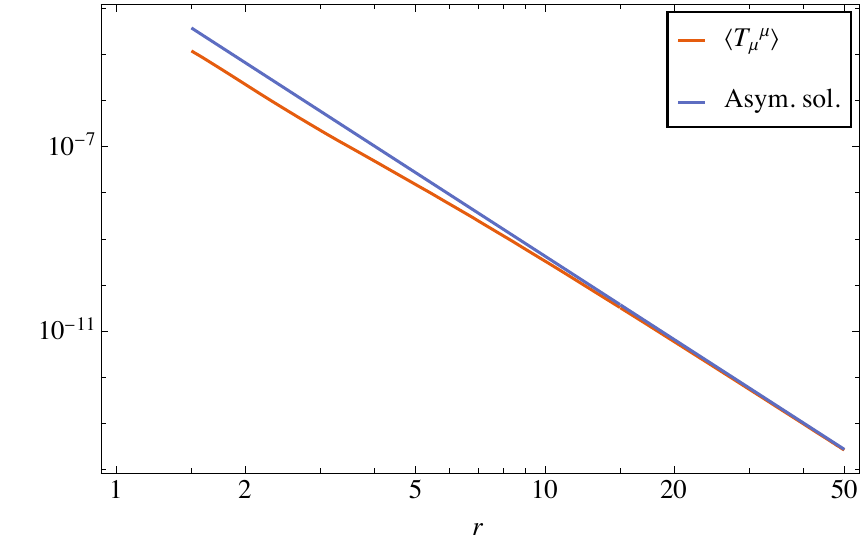}%
    \subcaption{$\ev{T\indices{_\mu^\mu}}$}%
    \label{fig:bd_SETtr_asym}%
  \end{minipage}%
  \caption{
  The asymptotic behaviors of each component and the trace of the RSET. The Asym. sol. denotes the analytical asymptotic solutions shown in Eq.~\eqref{eq:asymptotic_behavior_RSET}. 
  }
  \label{fig:bd_sol_asym}
\end{figure}%

The values of each component and trace of the RSET displayed in FIGs.~\ref{fig:bd_SET00}--\ref{fig:bd_SETtr} are regular everywhere, including the center as we have required. They vanish as $r\rightarrow\infty$, similar to those observed in the Boulware or Minkowski vacuum. As exhibited in \figref{fig:bd_sol_asym}, however, the decreasing power of RSET's components is as follows:
\begin{align}\label{eq:asymptotic_behavior_RSET}
    \ev{T\indices{_t^t}} \sim - \frac{q'(74 + 9c_{\rm H})}{51840 \pi^2 r^4}\hbar, \quad
    \ev{T\indices{_r^r}} \sim \frac{q'}{2880 \pi^2 M r^3} \hbar, \quad
    \ev{T\indices{_\theta^\theta}} \sim -\frac{q'}{5760 \pi^2 M r^3}\hbar, \quad
    \ev{T\indices{_\mu^\mu}} \sim \frac{M^2}{60 \pi^2 r^6}\hbar,
\end{align}
which is different from the behavior under the Boulware vacuum in Schwarzschild spacetime, where the decrease is as fast as $\sim M^2 / r^6$. This difference arises from the nonzero $\psi'$ or nonzero $q'$. The deviation from the Boulware vacuum also appears in the RSET profile.
\footnote{
In \cite{Balbinot:1999ri, Balbinot:1999vg}, it has already been pointed out that a deviation from the Boulware state arises when using the Riegert action \cite{Riegert:1984kt}, and the authors show that it appears from the subleading terms after reproducing the leading behavior. By contrast, it is impossible to reproduce even the leading term behavior due to the regularity at the center in our case. 
}

Here we comment on the Minkowski limit. In the $M\rightarrow 0$ limit, it can be observed that the parameters $\varphi_2$, $\psi_2$, $q$, and $q'$ tend towards zero. This is consistent with the expectation that trivial solutions are produced at the limit. It can be seen that the remaining parameters, $c_{\rm H}$, and $d_{\rm H}$, need not vanish since they appear with higher-order terms of compactness $M/r$ in Eq.~\eqref{eq:phi_4dSch_expansion}. These terms are naturally suppressed for the Minkowski limit.

The deviation from the Boulware vacuum stems from the nonconformal flatness of this four-dimensional spacetime~\eqref{eq:Bardeen_met} [or $F\ne 0$ in Eq.~\eqref{eq:Laction4d}, i.e. the breaking of the $\varphi$-shift symmetry]. In contrast to the two-dimensional case in \secref{sec:2D_SSS}, even horizonless spacetimes have different global structures from the Minkowski spacetime in the four-dimensional case. Thus similarly to the four-dimensional Schwarzschild case, this deviation from the conformal flatness forces the fields to diverge either at the center or at the (future and past) infinities. In the Schwarzschild spacetime, the Boulware vacuum is realized by imposing the singular behavior on the near-horizon region. On the other hand, in the case of the regular spacetime case, the central regularities of $\varphi, \psi$, and RSET lead to the loss of convergence of $\psi$ at infinity as a trade-off. 

\subsubsection{Conformally flat case}

Conversely, the central regularities and the Boulware vacuum (regularities at $r\rightarrow\infty$ in other words) are compatible if the spacetime has no singularity and is conformally flat since in this case $F=0$ and thus the $\varphi$-shift symmetry is respected. In such a spacetime with a metric
\begin{align}
    ds^2 = f\qty(-\dd{t}^2+\dd{r}^{2} +r^2 \dd{\Omega}^2),
\end{align}
the field equations \eqref{eq:EOM_phi4d} and \eqref{eq:EOM_psi4d} have analytic solutions
\begin{align}
    \varphi = \frac{c_{-1}}{r} + c_0 + c_1 r + c_2 r^2 + \log f(r), \quad
    \psi = \frac{d_{-1}}{r} + d_0 + d_1 r + d_2 r^2.
    \label{eq:sol_conformal4D}
\end{align}
As in the two-dimensional case, we can obtain the solutions with regularities at both $r \rightarrow 0$ and $r \rightarrow \infty$, i,e. Boulware vacuum with central regularities by properly choosing eight coefficients $\qty{c_{-1}, c_0, c_1, c_2, d_{-1}, d_0, d_1, d_2}$ if $f(r)$ is regular. Moreover, this conformal flat nature leads to the existence of another Noether current. Since the quadratic form of Weyl tensor $F$ in the field equation \eqref{eq:EOM_psi4d} vanish globally, it is shown that a current
\begin{align}
    J^{\prime \mu} 
    = \nabla^\mu \square \psi
    +2\left(R^{\mu\nu}-\frac{R}{3} g^{\mu\nu}\right) \nabla_\nu \psi,
\end{align}
is conserved. This corresponds to the additional symmetry of constant shift $\varphi \rightarrow \varphi + \sigma_0$ \cite{Mottola:2006ew}, which means the global Weyl symmetry. One can also define the conserved charge similarly to Eq.~\eqref{eq:charge_4D} as
\begin{align}
    \tilde{q}'= \frac{1}{4\pi} \int_{\partial\Sigma} J^{\prime\mu} \dd{\Sigma}_\mu.
    \label{eq:charge'_4D}
\end{align}
This charge is related to $d_1$ in Eq.~\eqref{eq:sol_conformal4D} and $c_1$ for $\tilde{q}$ or $q$ defined in Eq.~\eqref{eq:charge_4D}. The regularities of Eq.~\eqref{eq:sol_conformal4D} result that both charges must vanish in a conformally flat spacetime.

\section{Discussion and Conclusion \label{sec:discussion}}

In this work, we have investigated the possible vacuum state and the renormalized stress-energy tensors of massless quantum fields in two- and four-dimensional horizonless regular spacetimes with spherical symmetry and staticity. We employed the formalism established in \cite{Riegert:1984kt, Shapiro:1994ww, Balbinot:1999vg, Balbinot:1999ri, Mazur:2001aa, Mottola:2006ew} that introduces local auxiliary scalar fields. This formalism simplifies the evaluation of the RSETs into two steps: solving the field equations for the auxiliary fields and calculating those SETs as in classical field theory. While the inhomogeneous solutions of the scalar fields correspond to the local higher-curvature correction, the homogeneous parts represent the nonlocal nature of quantum states and RSETs, i.e. the vacuum choice and the global properties of the spacetime. Unlike formalisms with explicit regularization procedures, this method facilitates an extended discussion into more general spacetime cases. Furthermore, this formalism offers a systematic classification of possible vacuum states. This formulation is a useful one for understanding the effects of quantum fields on curved spacetime.

In this framework, the boundary conditions determine the vacuum state associated with it. For a two-dimensional spacetime without horizons, the asymptotic Minkowski condition (vanishing scalar fields and RSET) at infinity and their regularities at the center are compatible, leading to the Boulware vacuum, which is similar to the Schwarzschild case. However, we found this is not the case for the four-dimensional spacetime. The asymptotic Minkowski condition at infinity and the central regularities are incompatible. Therefore, the Boulware vacuum is not realized, but instead, a nontrivial vacuum with a linearly increasing auxiliary scalar field is realized when we impose the central regularities as boundary conditions.

This difference comes from the conformal nonflatness of the spherically symmetric spacetime that we have studied in four dimensions. The conformal flatness allows one to relate vacuum states in curved spacetime to those in flat spacetime. We have investigated the existence of the vacuum state associated with the Minkowski vacuum. In the two-dimensional case where the geometry is conformally flat, one can find the vacuum state that is associated with the Minkowski vacuum, with an additional inhomogeneous part. However, in the four-dimensional case where the conformal flatness is absent, such a vacuum state cannot be generally constructed. This results in the divergence of the auxiliary field either at the center or at infinity.

One may think that the discrepancy in the asymptotic behavior of the RSET at infinity~\eqref{eq:asymptotic_behavior_RSET} seems similar to the result in \cite{Balbinot:1999ri, Balbinot:1999vg}. In their analysis, the RSET obtained from the Riegert action \cite{Riegert:1984kt} can reproduce the leading behavior near the horizon and at infinity for the Boulware state, the Unruh state, and the Israel-Hartle-Hawking state, respectively. On the other hand, they show that subleading behavior cannot be reproduced in each state. This comes from the fact that Weyl invariant terms are usually neglected when calculating the RSET from the anomaly-induced effective action. Such circumstances apply to our case as well. 
In our case, we cannot reproduce even the leading behavior of the Boulware state since there are no degrees of freedom to do so. Nevertheless, our analysis cannot completely exclude the possibility that the Boulware state is actually consistent due to the disregarded Weyl invariant term. Precisely speaking, what we have shown is that one of the following three conditions is incompatible with the others; 
\begin{inparaenum}[(1)]
    \item regularities of the auxiliary scalar fields and RSET at $r=0$,
    \item Boulware state,
    \item the anomaly-induced effective 
    action \cite{Riegert:1984kt} or its local field formulation \cite{Riegert:1984kt, Shapiro:1994ww, Balbinot:1999vg, Balbinot:1999ri, Mazur:2001aa, Mottola:2006ew}.
\end{inparaenum}
Although further investigation is needed to identify which factor causes this incompatibility, our results imply that the choice of vacuum requires attention.

This work focused on the proper configuration of the static auxiliary scalar fields and their SETs in horizonless regular spacetimes. In principle, the RSET could be computed by summing over mode functions of quantum fields, followed by regularization and subtraction of divergences. In regular spacetime, the mode functions should satisfy the regularity at the center. If the central regularity affects the long-distance behavior of low-frequency modes and if their contributions to the RSET is non-negligible, then we might observe nontrivial asymptotic behavior of the RSET in the distant region, similarly to how the (static and dynamical) tidal Love number of a regular compact object depends on the internal structure of the object. Such a cross-check analysis based on mode decomposition will deepen our understanding of semiclassical gravity.

Moving forward, it would be important to consider the evaluation of backreaction to the geometry (semiclassical analysis) for a more realistic approach as \cite{Carballo-Rubio:2017tlh, Arrechea:2021xkp, Reyes:2023fde, Arrechea:2023oax}. Their primary claim about the existence of the ultracompact star resulting from the semiclassical effect can be reexamined to check the consistency with the global regularities of RSET implemented by proper vacuum choice. Furthermore, investigating the influence of this phenomenon on stellar compactness could be suggestive in determining the compatibility of beyond-standard models containing numerous copies of particle species \cite{Dvali:2009fw, Dvali:2009ne}. We leave these studies as future work.

Another potential area for future study involves discussing the semiclassical effects in collapsing or evaporating BH spacetime using the local field formulation. The Hawking radiation and its RSET in such spacetime have been central topics in QFT in curved spacetime \cite{Davies:1976ei, Unruh:1976db, Parentani:1994ij}. Particularly, a recent study by Mottola \textit{et al}. \cite{Mottola:2023jlo} focused on the two-dimensional case and suggested that various options of initial states, including nonvacuum ones, can be considered based on the local field formulation. This could lead to modification of the spectrum of the Hawking flux. It would also be interesting to extend this discussion to the four-dimensional case and find possible initial states consistent with the regularities in the Minkowski region before collapsing.

\begin{acknowledgements}
The authors are deeply grateful for the invaluable discussions and insightful comments by Julio Arrechia and Valeri Frolov. K.N. is grateful for the hospitality of YITP members during his visit as the atom-type fellow. This work is supported in part by Japan Society for the Promotion of Science (JSPS) KAKENHI Grants No.~JP23KJ1090 (K.N.), No.~JP23KJ1162 (K.O.), and No.~JP24K07017 (S.M.) as well as the World Premier International Research Center Initiative (WPI), MEXT, Japan (S.M.). 
\end{acknowledgements}

\appendix

\section{Comment on the horizonless Hayward-type spacetime \label{sec:hayward}}

As well as the Bardeen-type solution \cite{1968qtr..conf...87B, Carballo-Rubio:2022nuj} we used, another well-known example of regular spacetimes is the Hayward-type one \cite{Hayward:2005gi, Carballo-Rubio:2022nuj}, whose metric is 
\begin{align}
    \dd{s^2}
    =-f(r) \dd{t}^2+  f(r)^{-1} \dd{r}^2
    +r^2 \dd{\Omega}^2,
    \label{eq:heyward_met}
\end{align}
where
\begin{align}
    f(r)=1-\frac{2 m(r)}{r}, \quad m(r)=\frac{M r^3}{r^3 + R^3}.
\end{align}
This geometry also does or does not possess horizons depending on the mass $M$, while it is free from the central divergent singularity. 

However, it has a discontinuous behavior in its derivatives around the center; the curvature is found to be
\begin{align}
    \mathcal{R}(r) = \frac{-12MR^3(r^3-2R^3)}{(r^3+R^3)^3}
    = \frac{24 M}{R^3}-\frac{84 M r^3}{R^6}+\frac{180 M r^6}{R^9}+\order{r^7}.
\end{align}
The curvature is not a $C^3$-class function around the origin. Thus this geometry is not ``regular'' in this sense, and fields residing there can neither be. In fact, the auxiliary scalar field $\varphi$ with field equation \eqref{eq:EOM_phi4d} in this spacetime asymptotes
\begin{align}
    \varphi(r) = \frac{M\left(2 M+3 R^3 \varphi_2\right)}{5 R^6} r^4 + \frac{14M}{15R^6}r^5 
    + \frac{2 M^2\left(8 M+9 R^3 \varphi_2\right)}{21 R^9} r^6
    + \order{r^7},
\end{align}
[see Eq.~\eqref{eq:scsol_center} for $\varphi_2$] showing that it is not a $C^5$-class function around the center. 
This behavior leads to the RSET linear to $r$.
This feature stems from the geometry, and cannot be removed.  One should keep in mind when investigating regularities of the fields in such a spacetime, since it confuses us with the origin of the irregular behaviors.

\section{Junction conditions at discontinuous hypersurfaces \label{sec:JC}}

In general spacetime scenarios, the distribution of classical matter in the background may be discontinuous or exhibit a delta-function-like singular behavior, leading to a corresponding discontinuity or a delta-function-like divergence in the curvature. Hereafter, by discontinuity or a delta-function-like divergence, we mean that some physical or geometrical quantities (e.g. extrinsic curvature or curvature) change appreciably in timescales or length scales that are sufficiently longer than the cutoff of the theory under consideration but that are sufficiently shorter than any other physical scales of interest. In such instances, the field equations of the scalar fields in two-dimensional spacetime [Eq.~\eqref{eq:EOM_phi2d}] or in four-dimensional spacetime [Eqs.~\eqref{eq:EOM_phi4d} and \eqref{eq:EOM_psi4d}], exhibit discontinuities. As a result, we need to impose junction conditions at these discontinuous or singular hypersurfaces. 

In a two-dimensional spacetime case with a single scalar field that obeys Eq.~\eqref{eq:EOM_phi2d}, the first condition is provided from the conservation of the Noether current $J^{\mu}$~\eqref{eq:current_2D}. For example, one can consider a spacetime with an at most delta-function-like behavior of curvature on a singular hypersurface $r=r_s$. The conservation law should be still valid in a neighborhood of the singular hypersurface so that the integration across $r=r_s$ gives
\begin{align}
    \qty[n^{\mu} J_{\mu}] 
    = \qty[n^{\mu} \qty(\nabla_{\mu} \varphi + \Omega_{\mu})]
    \coloneqq \eval{n^{\mu} \qty(\nabla_{\mu} \varphi + \Omega_{\mu})}_{r \rightarrow r_s + 0}
    - \eval{n^{\mu} \qty(\nabla_{\mu} \varphi + \Omega_{\mu})}_{r \rightarrow r_s - 0}
    = 0,
\end{align}
where $n^\mu$ is a normal vector to the surface $r=r_s$. This condition fixes the jump in the field's first derivative in terms of that in geometry
\begin{align}
    \qty[n^{\mu} \nabla_{\mu} \varphi] 
    = -\qty[n^{\mu} \Omega_{\mu}].
    \label{eq:jc_2d_1}
\end{align}
Note that $\Omega^{\mu}$ contains up to first-order derivative of the metric, as noticed from the definition \eqref{eq:Omega_2D}. Hence, the above condition reduces to the continuity of $n^{\mu} \nabla_{\mu} \varphi$, if the discontinuity in curvature is step-function-like. Even if the curvature exhibits a delta-function-like behavior across the singular hypersurface, the jump in $n^{\mu} \nabla_{\mu} \varphi$ is step-function-like at most. In any case, the above condition \eqref{eq:jc_2d_1} requires that $\varphi$ itself must be continuous at the singular surface:
\begin{align}
    \qty[\varphi] = 0.
    \label{eq:jc_2d_2}
\end{align}
The conditions \eqref{eq:jc_2d_1} and \eqref{eq:jc_2d_2} are the appropriate junction conditions.

In four-dimensional spherically symmetric spacetime with a step-function-like discontinuity in curvature at $r=r_s$, one can find the junction conditions. 
The conservation law of the current $J^\mu$~\eqref{eq:current_4D} is still valid around $r=r_s$ irrelevant to the matter distribution as the two-dimensional case, so that
\begin{align}
    \qty[n^{\mu} J_{\mu}] = 0.
\end{align}
According to Eq.~\eqref{eq:current_4D}, this condition provides the junction condition of the third derivative of $\varphi$
\begin{align}
    \qty[n_{\mu} \nabla^\mu \left(\Box \varphi +\frac{1}{3} R\right)]  
    = - \qty[2n_{\mu}\left(R^{\mu\nu}-\frac{R}{3} g^{\mu\nu}\right) \nabla_\nu \varphi-\frac{1}{2} n_{\mu} \Omega^\mu].
    \label{eq:junction_4d_phi'''}
\end{align}
In the present case with a step-function-like discontinuity in curvature, the right-hand side is finite.
For another auxiliary field $\psi$, the condition is obtained by direct integration of the field equation \eqref{eq:EOM_psi4d}.
The condition \eqref{eq:junction_4d_phi'''} and the finiteness of the right-hand side of it imply that the second derivative of $\varphi$ must obey the following condition:
\begin{align}
    \qty[\Box \varphi] = -\qty[\frac{1}{3} R ].
\end{align}
Considering the at most step-function-like discontinuity in curvature, similar discussions lead to the continuity conditions for $\varphi$ and its first derivative;
\begin{align}
    &\qty[n_{\mu} \nabla^{\mu}\varphi] = 0, \\[5pt]
    &\qty[\varphi] = 0.
    \label{eq:jc_4d_4}
\end{align}
The conditions \eqref{eq:junction_4d_phi'''}--\eqref{eq:jc_4d_4} are the junction conditions for $\varphi$. One can find similar conditions for $\psi$, although we do not exhibit them here since they are not written in simple boundary value forms.

\bibliography{ref}

\begin{thebibliography}{71}%
\makeatletter
\providecommand \@ifxundefined [1]{%
 \@ifx{#1\undefined}
}%
\providecommand \@ifnum [1]{%
 \ifnum #1\expandafter \@firstoftwo
 \else \expandafter \@secondoftwo
 \fi
}%
\providecommand \@ifx [1]{%
 \ifx #1\expandafter \@firstoftwo
 \else \expandafter \@secondoftwo
 \fi
}%
\providecommand \natexlab [1]{#1}%
\providecommand \enquote  [1]{``#1''}%
\providecommand \bibnamefont  [1]{#1}%
\providecommand \bibfnamefont [1]{#1}%
\providecommand \citenamefont [1]{#1}%
\providecommand \href@noop [0]{\@secondoftwo}%
\providecommand \href [0]{\begingroup \@sanitize@url \@href}%
\providecommand \@href[1]{\@@startlink{#1}\@@href}%
\providecommand \@@href[1]{\endgroup#1\@@endlink}%
\providecommand \@sanitize@url [0]{\catcode `\\12\catcode `\$12\catcode `\&12\catcode `\#12\catcode `\^12\catcode `\_12\catcode `\%12\relax}%
\providecommand \@@startlink[1]{}%
\providecommand \@@endlink[0]{}%
\providecommand \url  [0]{\begingroup\@sanitize@url \@url }%
\providecommand \@url [1]{\endgroup\@href {#1}{\urlprefix }}%
\providecommand \urlprefix  [0]{URL }%
\providecommand \Eprint [0]{\href }%
\providecommand \doibase [0]{https://doi.org/}%
\providecommand \selectlanguage [0]{\@gobble}%
\providecommand \bibinfo  [0]{\@secondoftwo}%
\providecommand \bibfield  [0]{\@secondoftwo}%
\providecommand \translation [1]{[#1]}%
\providecommand \BibitemOpen [0]{}%
\providecommand \bibitemStop [0]{}%
\providecommand \bibitemNoStop [0]{.\EOS\space}%
\providecommand \EOS [0]{\spacefactor3000\relax}%
\providecommand \BibitemShut  [1]{\csname bibitem#1\endcsname}%
\let\auto@bib@innerbib\@empty
\bibitem [{\citenamefont {Capper}\ and\ \citenamefont {Duff}(1974)}]{Capper:1974ic}%
  \BibitemOpen
  \bibfield  {author} {\bibinfo {author} {\bibfnamefont {D.~M.}\ \bibnamefont {Capper}}\ and\ \bibinfo {author} {\bibfnamefont {M.~J.}\ \bibnamefont {Duff}},\ }\bibfield  {title} {\bibinfo {title} {{Trace anomalies in dimensional regularization}},\ }\href {https://doi.org/10.1007/BF02748300} {\bibfield  {journal} {\bibinfo  {journal} {Nuovo Cim. A}\ }\textbf {\bibinfo {volume} {23}},\ \bibinfo {pages} {173} (\bibinfo {year} {1974})}\BibitemShut {NoStop}%
\bibitem [{\citenamefont {Capper}\ and\ \citenamefont {Duff}(1975)}]{Capper:1975ig}%
  \BibitemOpen
  \bibfield  {author} {\bibinfo {author} {\bibfnamefont {D.~M.}\ \bibnamefont {Capper}}\ and\ \bibinfo {author} {\bibfnamefont {M.~J.}\ \bibnamefont {Duff}},\ }\bibfield  {title} {\bibinfo {title} {{Conformal Anomalies and the Renormalizability Problem in Quantum Gravity}},\ }\href {https://doi.org/10.1016/0375-9601(75)90030-4} {\bibfield  {journal} {\bibinfo  {journal} {Phys. Lett. A}\ }\textbf {\bibinfo {volume} {53}},\ \bibinfo {pages} {361} (\bibinfo {year} {1975})}\BibitemShut {NoStop}%
\bibitem [{\citenamefont {Deser}\ \emph {et~al.}(1976)\citenamefont {Deser}, \citenamefont {Duff},\ and\ \citenamefont {Isham}}]{Deser:1976yx}%
  \BibitemOpen
  \bibfield  {author} {\bibinfo {author} {\bibfnamefont {S.}~\bibnamefont {Deser}}, \bibinfo {author} {\bibfnamefont {M.~J.}\ \bibnamefont {Duff}},\ and\ \bibinfo {author} {\bibfnamefont {C.~J.}\ \bibnamefont {Isham}},\ }\bibfield  {title} {\bibinfo {title} {{Nonlocal Conformal Anomalies}},\ }\href {https://doi.org/10.1016/0550-3213(76)90480-6} {\bibfield  {journal} {\bibinfo  {journal} {Nucl. Phys. B}\ }\textbf {\bibinfo {volume} {111}},\ \bibinfo {pages} {45} (\bibinfo {year} {1976})}\BibitemShut {NoStop}%
\bibitem [{\citenamefont {Duff}(1977)}]{Duff:1977ay}%
  \BibitemOpen
  \bibfield  {author} {\bibinfo {author} {\bibfnamefont {M.~J.}\ \bibnamefont {Duff}},\ }\bibfield  {title} {\bibinfo {title} {{Observations on Conformal Anomalies}},\ }\href {https://doi.org/10.1016/0550-3213(77)90410-2} {\bibfield  {journal} {\bibinfo  {journal} {Nucl. Phys. B}\ }\textbf {\bibinfo {volume} {125}},\ \bibinfo {pages} {334} (\bibinfo {year} {1977})}\BibitemShut {NoStop}%
\bibitem [{\citenamefont {Birrell}\ and\ \citenamefont {Davies}(1984)}]{Birrell:1982ix}%
  \BibitemOpen
  \bibfield  {author} {\bibinfo {author} {\bibfnamefont {N.~D.}\ \bibnamefont {Birrell}}\ and\ \bibinfo {author} {\bibfnamefont {P.~C.~W.}\ \bibnamefont {Davies}},\ }\href {https://doi.org/10.1017/CBO9780511622632} {\emph {\bibinfo {title} {{Quantum Fields in Curved Space}}}},\ Cambridge Monographs on Mathematical Physics\ (\bibinfo  {publisher} {Cambridge Univ. Press},\ \bibinfo {address} {Cambridge, UK},\ \bibinfo {year} {1984})\BibitemShut {NoStop}%
\bibitem [{\citenamefont {Parker}\ and\ \citenamefont {Toms}(2009)}]{Parker:2009uva}%
  \BibitemOpen
  \bibfield  {author} {\bibinfo {author} {\bibfnamefont {L.~E.}\ \bibnamefont {Parker}}\ and\ \bibinfo {author} {\bibfnamefont {D.}~\bibnamefont {Toms}},\ }\href {https://doi.org/10.1017/CBO9780511813924} {\emph {\bibinfo {title} {{Quantum Field Theory in Curved Spacetime}: {Quantized Field and Gravity}}}},\ Cambridge Monographs on Mathematical Physics\ (\bibinfo  {publisher} {Cambridge University Press},\ \bibinfo {year} {2009})\BibitemShut {NoStop}%
\bibitem [{\citenamefont {Hu}\ and\ \citenamefont {Verdaguer}(2020)}]{Hu:2020luk}%
  \BibitemOpen
  \bibfield  {author} {\bibinfo {author} {\bibfnamefont {B.-L.~B.}\ \bibnamefont {Hu}}\ and\ \bibinfo {author} {\bibfnamefont {E.}~\bibnamefont {Verdaguer}},\ }\href {https://doi.org/10.1017/9780511667497} {\emph {\bibinfo {title} {{Semiclassical and Stochastic Gravity}: {Quantum Field Effects on Curved Spacetime}}}},\ Cambridge Monographs on Mathematical Physics\ (\bibinfo  {publisher} {Cambridge University Press},\ \bibinfo {address} {Cambridge},\ \bibinfo {year} {2020})\BibitemShut {NoStop}%
\bibitem [{\citenamefont {Duff}(1994)}]{Duff_1994}%
  \BibitemOpen
  \bibfield  {author} {\bibinfo {author} {\bibfnamefont {M.~J.}\ \bibnamefont {Duff}},\ }\bibfield  {title} {\bibinfo {title} {Twenty years of the weyl anomaly},\ }\href {https://doi.org/10.1088/0264-9381/11/6/004} {\bibfield  {journal} {\bibinfo  {journal} {Classical and Quantum Gravity}\ }\textbf {\bibinfo {volume} {11}},\ \bibinfo {pages} {1387} (\bibinfo {year} {1994})}\BibitemShut {NoStop}%
\bibitem [{\citenamefont {Hawking}(1974)}]{Hawking:1974rv}%
  \BibitemOpen
  \bibfield  {author} {\bibinfo {author} {\bibfnamefont {S.~W.}\ \bibnamefont {Hawking}},\ }\bibfield  {title} {\bibinfo {title} {{Black hole explosions}},\ }\href {https://doi.org/10.1038/248030a0} {\bibfield  {journal} {\bibinfo  {journal} {Nature}\ }\textbf {\bibinfo {volume} {248}},\ \bibinfo {pages} {30} (\bibinfo {year} {1974})}\BibitemShut {NoStop}%
\bibitem [{\citenamefont {Davies}(1975)}]{Davies:1974th}%
  \BibitemOpen
  \bibfield  {author} {\bibinfo {author} {\bibfnamefont {P.~C.~W.}\ \bibnamefont {Davies}},\ }\bibfield  {title} {\bibinfo {title} {{Scalar particle production in Schwarzschild and Rindler metrics}},\ }\href {https://doi.org/10.1088/0305-4470/8/4/022} {\bibfield  {journal} {\bibinfo  {journal} {J. Phys. A}\ }\textbf {\bibinfo {volume} {8}},\ \bibinfo {pages} {609} (\bibinfo {year} {1975})}\BibitemShut {NoStop}%
\bibitem [{\citenamefont {Hawking}(1975)}]{Hawking:1975vcx}%
  \BibitemOpen
  \bibfield  {author} {\bibinfo {author} {\bibfnamefont {S.~W.}\ \bibnamefont {Hawking}},\ }\bibfield  {title} {\bibinfo {title} {{Particle Creation by Black Holes}},\ }\href {https://doi.org/10.1007/BF02345020} {\bibfield  {journal} {\bibinfo  {journal} {Commun. Math. Phys.}\ }\textbf {\bibinfo {volume} {43}},\ \bibinfo {pages} {199} (\bibinfo {year} {1975})},\ \bibinfo {note} {[Erratum: Commun.Math.Phys. 46, 206 (1976)]}\BibitemShut {NoStop}%
\bibitem [{\citenamefont {Wald}(1975)}]{Wald:1975kc}%
  \BibitemOpen
  \bibfield  {author} {\bibinfo {author} {\bibfnamefont {R.~M.}\ \bibnamefont {Wald}},\ }\bibfield  {title} {\bibinfo {title} {{On Particle Creation by Black Holes}},\ }\href {https://doi.org/10.1007/BF01609863} {\bibfield  {journal} {\bibinfo  {journal} {Commun. Math. Phys.}\ }\textbf {\bibinfo {volume} {45}},\ \bibinfo {pages} {9} (\bibinfo {year} {1975})}\BibitemShut {NoStop}%
\bibitem [{\citenamefont {Parker}(1975)}]{Parker:1975jm}%
  \BibitemOpen
  \bibfield  {author} {\bibinfo {author} {\bibfnamefont {L.}~\bibnamefont {Parker}},\ }\bibfield  {title} {\bibinfo {title} {{Probability Distribution of Particles Created by a Black Hole}},\ }\href {https://doi.org/10.1103/PhysRevD.12.1519} {\bibfield  {journal} {\bibinfo  {journal} {Phys. Rev. D}\ }\textbf {\bibinfo {volume} {12}},\ \bibinfo {pages} {1519} (\bibinfo {year} {1975})}\BibitemShut {NoStop}%
\bibitem [{\citenamefont {Parker}(1968)}]{Parker:1968mv}%
  \BibitemOpen
  \bibfield  {author} {\bibinfo {author} {\bibfnamefont {L.}~\bibnamefont {Parker}},\ }\bibfield  {title} {\bibinfo {title} {{Particle creation in expanding universes}},\ }\href {https://doi.org/10.1103/PhysRevLett.21.562} {\bibfield  {journal} {\bibinfo  {journal} {Phys. Rev. Lett.}\ }\textbf {\bibinfo {volume} {21}},\ \bibinfo {pages} {562} (\bibinfo {year} {1968})}\BibitemShut {NoStop}%
\bibitem [{\citenamefont {Parker}(1969)}]{Parker:1969au}%
  \BibitemOpen
  \bibfield  {author} {\bibinfo {author} {\bibfnamefont {L.}~\bibnamefont {Parker}},\ }\bibfield  {title} {\bibinfo {title} {{Quantized fields and particle creation in expanding universes. 1.}},\ }\href {https://doi.org/10.1103/PhysRev.183.1057} {\bibfield  {journal} {\bibinfo  {journal} {Phys. Rev.}\ }\textbf {\bibinfo {volume} {183}},\ \bibinfo {pages} {1057} (\bibinfo {year} {1969})}\BibitemShut {NoStop}%
\bibitem [{\citenamefont {Sexl}\ and\ \citenamefont {Urbantke}(1969)}]{Sexl:1969ix}%
  \BibitemOpen
  \bibfield  {author} {\bibinfo {author} {\bibfnamefont {R.~U.}\ \bibnamefont {Sexl}}\ and\ \bibinfo {author} {\bibfnamefont {H.~K.}\ \bibnamefont {Urbantke}},\ }\bibfield  {title} {\bibinfo {title} {{Production of particles by gravitational fields}},\ }\href {https://doi.org/10.1103/PhysRev.179.1247} {\bibfield  {journal} {\bibinfo  {journal} {Phys. Rev.}\ }\textbf {\bibinfo {volume} {179}},\ \bibinfo {pages} {1247} (\bibinfo {year} {1969})}\BibitemShut {NoStop}%
\bibitem [{\citenamefont {Zeldovich}(1970)}]{Zeldovich:1970si}%
  \BibitemOpen
  \bibfield  {author} {\bibinfo {author} {\bibfnamefont {Y.~B.}\ \bibnamefont {Zeldovich}},\ }\bibfield  {title} {\bibinfo {title} {{Particle production in cosmology}},\ }\href@noop {} {\bibfield  {journal} {\bibinfo  {journal} {Pisma Zh. Eksp. Teor. Fiz.}\ }\textbf {\bibinfo {volume} {12}},\ \bibinfo {pages} {443} (\bibinfo {year} {1970})}\BibitemShut {NoStop}%
\bibitem [{\citenamefont {Wald}(1978)}]{Wald:1978ce}%
  \BibitemOpen
  \bibfield  {author} {\bibinfo {author} {\bibfnamefont {R.~M.}\ \bibnamefont {Wald}},\ }\bibfield  {title} {\bibinfo {title} {{Axiomatic Renormalization of the Stress Tensor of a Conformally Invariant Field in Conformally Flat Space-Times}},\ }\href {https://doi.org/10.1016/0003-4916(78)90040-4} {\bibfield  {journal} {\bibinfo  {journal} {Annals Phys.}\ }\textbf {\bibinfo {volume} {110}},\ \bibinfo {pages} {472} (\bibinfo {year} {1978})}\BibitemShut {NoStop}%
\bibitem [{\citenamefont {{Davies}}\ and\ \citenamefont {{Fulling}}(1977)}]{1977RSPSA.354...59D}%
  \BibitemOpen
  \bibfield  {author} {\bibinfo {author} {\bibfnamefont {P.~C.~W.}\ \bibnamefont {{Davies}}}\ and\ \bibinfo {author} {\bibfnamefont {S.~A.}\ \bibnamefont {{Fulling}}},\ }\bibfield  {title} {\bibinfo {title} {{Quantum vacuum energy in two dimensional space-times}},\ }\href {https://doi.org/10.1098/rspa.1977.0056} {\bibfield  {journal} {\bibinfo  {journal} {Proc. R. Soc. Lond. A}\ }\textbf {\bibinfo {volume} {354}},\ \bibinfo {pages} {59} (\bibinfo {year} {1977})}\BibitemShut {NoStop}%
\bibitem [{\citenamefont {Fulling}(1977)}]{Fulling:1977jm}%
  \BibitemOpen
  \bibfield  {author} {\bibinfo {author} {\bibfnamefont {S.~A.}\ \bibnamefont {Fulling}},\ }\bibfield  {title} {\bibinfo {title} {{Radiation and Vacuum Polarization Near a Black Hole}},\ }\href {https://doi.org/10.1103/PhysRevD.15.2411} {\bibfield  {journal} {\bibinfo  {journal} {Phys. Rev. D}\ }\textbf {\bibinfo {volume} {15}},\ \bibinfo {pages} {2411} (\bibinfo {year} {1977})}\BibitemShut {NoStop}%
\bibitem [{\citenamefont {Christensen}\ and\ \citenamefont {Fulling}(1977)}]{Christensen:1977jc}%
  \BibitemOpen
  \bibfield  {author} {\bibinfo {author} {\bibfnamefont {S.~M.}\ \bibnamefont {Christensen}}\ and\ \bibinfo {author} {\bibfnamefont {S.~A.}\ \bibnamefont {Fulling}},\ }\bibfield  {title} {\bibinfo {title} {{Trace Anomalies and the Hawking Effect}},\ }\href {https://doi.org/10.1103/PhysRevD.15.2088} {\bibfield  {journal} {\bibinfo  {journal} {Phys. Rev. D}\ }\textbf {\bibinfo {volume} {15}},\ \bibinfo {pages} {2088} (\bibinfo {year} {1977})}\BibitemShut {NoStop}%
\bibitem [{\citenamefont {Candelas}(1980)}]{Candelas:1980zt}%
  \BibitemOpen
  \bibfield  {author} {\bibinfo {author} {\bibfnamefont {P.}~\bibnamefont {Candelas}},\ }\bibfield  {title} {\bibinfo {title} {{Vacuum Polarization in Schwarzschild Space-Time}},\ }\href {https://doi.org/10.1103/PhysRevD.21.2185} {\bibfield  {journal} {\bibinfo  {journal} {Phys. Rev. D}\ }\textbf {\bibinfo {volume} {21}},\ \bibinfo {pages} {2185} (\bibinfo {year} {1980})}\BibitemShut {NoStop}%
\bibitem [{\citenamefont {Frolov}\ and\ \citenamefont {Zelnikov}(1987)}]{Frolov:1987gw}%
  \BibitemOpen
  \bibfield  {author} {\bibinfo {author} {\bibfnamefont {V.~P.}\ \bibnamefont {Frolov}}\ and\ \bibinfo {author} {\bibfnamefont {A.~I.}\ \bibnamefont {Zelnikov}},\ }\bibfield  {title} {\bibinfo {title} {{Killing Approximation for Vacuum and Thermal Stress - Energy Tensor in Static Space-times}},\ }\href {https://doi.org/10.1103/PhysRevD.35.3031} {\bibfield  {journal} {\bibinfo  {journal} {Phys. Rev. D}\ }\textbf {\bibinfo {volume} {35}},\ \bibinfo {pages} {3031} (\bibinfo {year} {1987})}\BibitemShut {NoStop}%
\bibitem [{\citenamefont {Anderson}\ \emph {et~al.}(1995)\citenamefont {Anderson}, \citenamefont {Hiscock},\ and\ \citenamefont {Samuel}}]{Anderson:1994hg}%
  \BibitemOpen
  \bibfield  {author} {\bibinfo {author} {\bibfnamefont {P.~R.}\ \bibnamefont {Anderson}}, \bibinfo {author} {\bibfnamefont {W.~A.}\ \bibnamefont {Hiscock}},\ and\ \bibinfo {author} {\bibfnamefont {D.~A.}\ \bibnamefont {Samuel}},\ }\bibfield  {title} {\bibinfo {title} {{Stress - energy tensor of quantized scalar fields in static spherically symmetric space-times}},\ }\href {https://doi.org/10.1103/PhysRevD.51.4337} {\bibfield  {journal} {\bibinfo  {journal} {Phys. Rev. D}\ }\textbf {\bibinfo {volume} {51}},\ \bibinfo {pages} {4337} (\bibinfo {year} {1995})}\BibitemShut {NoStop}%
\bibitem [{\citenamefont {Bunch}\ and\ \citenamefont {Davies}(1977)}]{Bunch:1977sc}%
  \BibitemOpen
  \bibfield  {author} {\bibinfo {author} {\bibfnamefont {T.}~\bibnamefont {Bunch}}\ and\ \bibinfo {author} {\bibfnamefont {P.}~\bibnamefont {Davies}},\ }\bibfield  {title} {\bibinfo {title} {Stress tensor and conformal anomalies for massless fields in a robertson-walker universe},\ }\href {https://doi.org/10.1098/rspa.1977.0151} {\bibfield  {journal} {\bibinfo  {journal} {Proc. R. Soc. Lond. A}\ }\textbf {\bibinfo {volume} {356}},\ \bibinfo {pages} {569} (\bibinfo {year} {1977})}\BibitemShut {NoStop}%
\bibitem [{\citenamefont {Davies}\ \emph {et~al.}(1977)\citenamefont {Davies}, \citenamefont {Fulling}, \citenamefont {Christensen},\ and\ \citenamefont {Bunch}}]{Davies:1977ze}%
  \BibitemOpen
  \bibfield  {author} {\bibinfo {author} {\bibfnamefont {P.~C.~W.}\ \bibnamefont {Davies}}, \bibinfo {author} {\bibfnamefont {S.~A.}\ \bibnamefont {Fulling}}, \bibinfo {author} {\bibfnamefont {S.~M.}\ \bibnamefont {Christensen}},\ and\ \bibinfo {author} {\bibfnamefont {T.~S.}\ \bibnamefont {Bunch}},\ }\bibfield  {title} {\bibinfo {title} {{Energy Momentum Tensor of a Massless Scalar Quantum Field in a Robertson-Walker Universe}},\ }\href {https://doi.org/10.1016/0003-4916(77)90167-1} {\bibfield  {journal} {\bibinfo  {journal} {Annals Phys.}\ }\textbf {\bibinfo {volume} {109}},\ \bibinfo {pages} {108} (\bibinfo {year} {1977})}\BibitemShut {NoStop}%
\bibitem [{\citenamefont {Page}(1982)}]{Page:1982fm}%
  \BibitemOpen
  \bibfield  {author} {\bibinfo {author} {\bibfnamefont {D.~N.}\ \bibnamefont {Page}},\ }\bibfield  {title} {\bibinfo {title} {{Thermal Stress Tensors in Static Einstein Spaces}},\ }\href {https://doi.org/10.1103/PhysRevD.25.1499} {\bibfield  {journal} {\bibinfo  {journal} {Phys. Rev. D}\ }\textbf {\bibinfo {volume} {25}},\ \bibinfo {pages} {1499} (\bibinfo {year} {1982})}\BibitemShut {NoStop}%
\bibitem [{\citenamefont {Brown}\ \emph {et~al.}(1986)\citenamefont {Brown}, \citenamefont {Ottewill},\ and\ \citenamefont {Page}}]{Brown:1986jy}%
  \BibitemOpen
  \bibfield  {author} {\bibinfo {author} {\bibfnamefont {M.~R.}\ \bibnamefont {Brown}}, \bibinfo {author} {\bibfnamefont {A.~C.}\ \bibnamefont {Ottewill}},\ and\ \bibinfo {author} {\bibfnamefont {D.~N.}\ \bibnamefont {Page}},\ }\bibfield  {title} {\bibinfo {title} {{Conformally Invariant Quantum Field Theory in Static Einstein Space-times}},\ }\href {https://doi.org/10.1103/PhysRevD.33.2840} {\bibfield  {journal} {\bibinfo  {journal} {Phys. Rev. D}\ }\textbf {\bibinfo {volume} {33}},\ \bibinfo {pages} {2840} (\bibinfo {year} {1986})}\BibitemShut {NoStop}%
\bibitem [{\citenamefont {Parker}(1979)}]{Parker:1978gh}%
  \BibitemOpen
  \bibfield  {author} {\bibinfo {author} {\bibfnamefont {L.}~\bibnamefont {Parker}},\ }\bibfield  {title} {\bibinfo {title} {{Aspects of Quantum Field Theory in Curved Space-Time: Effective Action and Energy Momentum Tensor}},\ }\href@noop {} {\bibfield  {journal} {\bibinfo  {journal} {NATO Sci. Ser. B}\ }\textbf {\bibinfo {volume} {44}},\ \bibinfo {pages} {219} (\bibinfo {year} {1979})}\BibitemShut {NoStop}%
\bibitem [{\citenamefont {{Lukash}}\ and\ \citenamefont {{Starobinski{\v{i}}}}(1974)}]{1974JETP...39..742L}%
  \BibitemOpen
  \bibfield  {author} {\bibinfo {author} {\bibfnamefont {V.~N.}\ \bibnamefont {{Lukash}}}\ and\ \bibinfo {author} {\bibfnamefont {A.~A.}\ \bibnamefont {{Starobinski{\v{i}}}}},\ }\bibfield  {title} {\bibinfo {title} {{The isotropization of the cosmological expansion owing to particle production}},\ }\href@noop {} {\bibfield  {journal} {\bibinfo  {journal} {Soviet Journal of Experimental and Theoretical Physics}\ }\textbf {\bibinfo {volume} {39}},\ \bibinfo {pages} {742} (\bibinfo {year} {1974})}\BibitemShut {NoStop}%
\bibitem [{\citenamefont {Grishchuk}(1977)}]{Grishchuk:1977zz}%
  \BibitemOpen
  \bibfield  {author} {\bibinfo {author} {\bibfnamefont {L.~P.}\ \bibnamefont {Grishchuk}},\ }\bibfield  {title} {\bibinfo {title} {{Graviton Creation in the Early Universe}},\ }\href {https://doi.org/10.1111/j.1749-6632.1977.tb37064.x} {\bibfield  {journal} {\bibinfo  {journal} {Annals N. Y. Acad. Sci.}\ }\textbf {\bibinfo {volume} {302}},\ \bibinfo {pages} {439} (\bibinfo {year} {1977})}\BibitemShut {NoStop}%
\bibitem [{\citenamefont {Guth}(1981)}]{Guth:1980zm}%
  \BibitemOpen
  \bibfield  {author} {\bibinfo {author} {\bibfnamefont {A.~H.}\ \bibnamefont {Guth}},\ }\bibfield  {title} {\bibinfo {title} {{The Inflationary Universe: A Possible Solution to the Horizon and Flatness Problems}},\ }\href {https://doi.org/10.1103/PhysRevD.23.347} {\bibfield  {journal} {\bibinfo  {journal} {Phys. Rev. D}\ }\textbf {\bibinfo {volume} {23}},\ \bibinfo {pages} {347} (\bibinfo {year} {1981})}\BibitemShut {NoStop}%
\bibitem [{\citenamefont {Hajicek}\ and\ \citenamefont {Israel}(1980)}]{HAJICEK19809}%
  \BibitemOpen
  \bibfield  {author} {\bibinfo {author} {\bibfnamefont {P.}~\bibnamefont {Hajicek}}\ and\ \bibinfo {author} {\bibfnamefont {W.}~\bibnamefont {Israel}},\ }\bibfield  {title} {\bibinfo {title} {What, no black hole evaporation?},\ }\href {https://doi.org/https://doi.org/10.1016/0375-9601(80)90439-9} {\bibfield  {journal} {\bibinfo  {journal} {Phys. Lett. A}\ }\textbf {\bibinfo {volume} {80}},\ \bibinfo {pages} {9} (\bibinfo {year} {1980})}\BibitemShut {NoStop}%
\bibitem [{\citenamefont {Bardeen}(1981)}]{Bardeen:1981zz}%
  \BibitemOpen
  \bibfield  {author} {\bibinfo {author} {\bibfnamefont {J.~M.}\ \bibnamefont {Bardeen}},\ }\bibfield  {title} {\bibinfo {title} {{Black Holes Do Evaporate Thermally}},\ }\href {https://doi.org/10.1103/PhysRevLett.46.382} {\bibfield  {journal} {\bibinfo  {journal} {Phys. Rev. Lett.}\ }\textbf {\bibinfo {volume} {46}},\ \bibinfo {pages} {382} (\bibinfo {year} {1981})}\BibitemShut {NoStop}%
\bibitem [{\citenamefont {Israel}(1976)}]{Israel:1976ur}%
  \BibitemOpen
  \bibfield  {author} {\bibinfo {author} {\bibfnamefont {W.}~\bibnamefont {Israel}},\ }\bibfield  {title} {\bibinfo {title} {{Thermo field dynamics of black holes}},\ }\href {https://doi.org/10.1016/0375-9601(76)90178-X} {\bibfield  {journal} {\bibinfo  {journal} {Phys. Lett. A}\ }\textbf {\bibinfo {volume} {57}},\ \bibinfo {pages} {107} (\bibinfo {year} {1976})}\BibitemShut {NoStop}%
\bibitem [{\citenamefont {Hartle}\ and\ \citenamefont {Hawking}(1976)}]{Hartle:1976tp}%
  \BibitemOpen
  \bibfield  {author} {\bibinfo {author} {\bibfnamefont {J.~B.}\ \bibnamefont {Hartle}}\ and\ \bibinfo {author} {\bibfnamefont {S.~W.}\ \bibnamefont {Hawking}},\ }\bibfield  {title} {\bibinfo {title} {{Path Integral Derivation of Black Hole Radiance}},\ }\href {https://doi.org/10.1103/PhysRevD.13.2188} {\bibfield  {journal} {\bibinfo  {journal} {Phys. Rev. D}\ }\textbf {\bibinfo {volume} {13}},\ \bibinfo {pages} {2188} (\bibinfo {year} {1976})}\BibitemShut {NoStop}%
\bibitem [{\citenamefont {Unruh}(1976)}]{Unruh:1976db}%
  \BibitemOpen
  \bibfield  {author} {\bibinfo {author} {\bibfnamefont {W.~G.}\ \bibnamefont {Unruh}},\ }\bibfield  {title} {\bibinfo {title} {{Notes on black hole evaporation}},\ }\href {https://doi.org/10.1103/PhysRevD.14.870} {\bibfield  {journal} {\bibinfo  {journal} {Phys. Rev. D}\ }\textbf {\bibinfo {volume} {14}},\ \bibinfo {pages} {870} (\bibinfo {year} {1976})}\BibitemShut {NoStop}%
\bibitem [{\citenamefont {Boulware}(1975)}]{Boulware:1974dm}%
  \BibitemOpen
  \bibfield  {author} {\bibinfo {author} {\bibfnamefont {D.~G.}\ \bibnamefont {Boulware}},\ }\bibfield  {title} {\bibinfo {title} {{Quantum Field Theory in Schwarzschild and Rindler Spaces}},\ }\href {https://doi.org/10.1103/PhysRevD.11.1404} {\bibfield  {journal} {\bibinfo  {journal} {Phys. Rev. D}\ }\textbf {\bibinfo {volume} {11}},\ \bibinfo {pages} {1404} (\bibinfo {year} {1975})}\BibitemShut {NoStop}%
\bibitem [{\citenamefont {Boulware}(1976)}]{Boulware:1975fe}%
  \BibitemOpen
  \bibfield  {author} {\bibinfo {author} {\bibfnamefont {D.~G.}\ \bibnamefont {Boulware}},\ }\bibfield  {title} {\bibinfo {title} {{Hawking Radiation and Thin Shells}},\ }\href {https://doi.org/10.1103/PhysRevD.13.2169} {\bibfield  {journal} {\bibinfo  {journal} {Phys. Rev. D}\ }\textbf {\bibinfo {volume} {13}},\ \bibinfo {pages} {2169} (\bibinfo {year} {1976})}\BibitemShut {NoStop}%
\bibitem [{\citenamefont {{Bardeen}}(1968)}]{1968qtr..conf...87B}%
  \BibitemOpen
  \bibfield  {author} {\bibinfo {author} {\bibfnamefont {J.}~\bibnamefont {{Bardeen}}},\ }\bibfield  {title} {\bibinfo {title} {{Non-singular general relativistic gravitational collapse}},\ }in\ \href@noop {} {\emph {\bibinfo {booktitle} {Proceedings of the 5th International Conference on Gravitation and the Theory of Relativity}}}\ (\bibinfo {year} {1968})\ p.~\bibinfo {pages} {87}\BibitemShut {NoStop}%
\bibitem [{\citenamefont {Carballo-Rubio}\ \emph {et~al.}(2023)\citenamefont {Carballo-Rubio}, \citenamefont {Di~Filippo}, \citenamefont {Liberati},\ and\ \citenamefont {Visser}}]{Carballo-Rubio:2022nuj}%
  \BibitemOpen
  \bibfield  {author} {\bibinfo {author} {\bibfnamefont {R.}~\bibnamefont {Carballo-Rubio}}, \bibinfo {author} {\bibfnamefont {F.}~\bibnamefont {Di~Filippo}}, \bibinfo {author} {\bibfnamefont {S.}~\bibnamefont {Liberati}},\ and\ \bibinfo {author} {\bibfnamefont {M.}~\bibnamefont {Visser}},\ }\bibfield  {title} {\bibinfo {title} {{A connection between regular black holes and horizonless ultracompact stars}},\ }\href {https://doi.org/10.1007/JHEP08(2023)046} {\bibfield  {journal} {\bibinfo  {journal} {JHEP}\ }\textbf {\bibinfo {volume} {08}},\ \bibinfo {pages} {046}},\ \Eprint {https://arxiv.org/abs/2211.05817} {arXiv:2211.05817 [gr-qc]} \BibitemShut {NoStop}%
\bibitem [{\citenamefont {Carballo-Rubio}(2018)}]{Carballo-Rubio:2017tlh}%
  \BibitemOpen
  \bibfield  {author} {\bibinfo {author} {\bibfnamefont {R.}~\bibnamefont {Carballo-Rubio}},\ }\bibfield  {title} {\bibinfo {title} {{Stellar equilibrium in semiclassical gravity}},\ }\href {https://doi.org/10.1103/PhysRevLett.120.061102} {\bibfield  {journal} {\bibinfo  {journal} {Phys. Rev. Lett.}\ }\textbf {\bibinfo {volume} {120}},\ \bibinfo {pages} {061102} (\bibinfo {year} {2018})},\ \Eprint {https://arxiv.org/abs/1706.05379} {arXiv:1706.05379 [gr-qc]} \BibitemShut {NoStop}%
\bibitem [{\citenamefont {Arrechea}\ \emph {et~al.}(2022)\citenamefont {Arrechea}, \citenamefont {Barcel\'o}, \citenamefont {Carballo-Rubio},\ and\ \citenamefont {Garay}}]{Arrechea:2021xkp}%
  \BibitemOpen
  \bibfield  {author} {\bibinfo {author} {\bibfnamefont {J.}~\bibnamefont {Arrechea}}, \bibinfo {author} {\bibfnamefont {C.}~\bibnamefont {Barcel\'o}}, \bibinfo {author} {\bibfnamefont {R.}~\bibnamefont {Carballo-Rubio}},\ and\ \bibinfo {author} {\bibfnamefont {L.~J.}\ \bibnamefont {Garay}},\ }\bibfield  {title} {\bibinfo {title} {{Semiclassical relativistic stars}},\ }\href {https://doi.org/10.1038/s41598-022-19836-8} {\bibfield  {journal} {\bibinfo  {journal} {Sci. Rep.}\ }\textbf {\bibinfo {volume} {12}},\ \bibinfo {pages} {15958} (\bibinfo {year} {2022})},\ \Eprint {https://arxiv.org/abs/2110.15808} {arXiv:2110.15808 [gr-qc]} \BibitemShut {NoStop}%
\bibitem [{\citenamefont {Reyes}\ and\ \citenamefont {Tomaselli}(2023)}]{Reyes:2023fde}%
  \BibitemOpen
  \bibfield  {author} {\bibinfo {author} {\bibfnamefont {I.~A.}\ \bibnamefont {Reyes}}\ and\ \bibinfo {author} {\bibfnamefont {G.~M.}\ \bibnamefont {Tomaselli}},\ }\bibfield  {title} {\bibinfo {title} {{Quantum field theory on compact stars near the Buchdahl limit}},\ }\href {https://doi.org/10.1103/PhysRevD.108.065006} {\bibfield  {journal} {\bibinfo  {journal} {Phys. Rev. D}\ }\textbf {\bibinfo {volume} {108}},\ \bibinfo {pages} {065006} (\bibinfo {year} {2023})},\ \Eprint {https://arxiv.org/abs/2301.00826} {arXiv:2301.00826 [gr-qc]} \BibitemShut {NoStop}%
\bibitem [{\citenamefont {Arrechea}\ \emph {et~al.}(2024)\citenamefont {Arrechea}, \citenamefont {Barcel\'o}, \citenamefont {Carballo-Rubio},\ and\ \citenamefont {Garay}}]{Arrechea:2023oax}%
  \BibitemOpen
  \bibfield  {author} {\bibinfo {author} {\bibfnamefont {J.}~\bibnamefont {Arrechea}}, \bibinfo {author} {\bibfnamefont {C.}~\bibnamefont {Barcel\'o}}, \bibinfo {author} {\bibfnamefont {R.}~\bibnamefont {Carballo-Rubio}},\ and\ \bibinfo {author} {\bibfnamefont {L.~J.}\ \bibnamefont {Garay}},\ }\bibfield  {title} {\bibinfo {title} {{Ultracompact horizonless objects in order-reduced semiclassical gravity}},\ }\href {https://doi.org/10.1103/PhysRevD.109.104056} {\bibfield  {journal} {\bibinfo  {journal} {Phys. Rev. D}\ }\textbf {\bibinfo {volume} {109}},\ \bibinfo {pages} {104056} (\bibinfo {year} {2024})},\ \Eprint {https://arxiv.org/abs/2310.12668} {arXiv:2310.12668 [gr-qc]} \BibitemShut {NoStop}%
\bibitem [{\citenamefont {Flanagan}\ and\ \citenamefont {Hinderer}(2008)}]{Flanagan:2007ix}%
  \BibitemOpen
  \bibfield  {author} {\bibinfo {author} {\bibfnamefont {E.~E.}\ \bibnamefont {Flanagan}}\ and\ \bibinfo {author} {\bibfnamefont {T.}~\bibnamefont {Hinderer}},\ }\bibfield  {title} {\bibinfo {title} {{Constraining neutron star tidal Love numbers with gravitational wave detectors}},\ }\href {https://doi.org/10.1103/PhysRevD.77.021502} {\bibfield  {journal} {\bibinfo  {journal} {Phys. Rev. D}\ }\textbf {\bibinfo {volume} {77}},\ \bibinfo {pages} {021502} (\bibinfo {year} {2008})},\ \Eprint {https://arxiv.org/abs/0709.1915} {arXiv:0709.1915 [astro-ph]} \BibitemShut {NoStop}%
\bibitem [{\citenamefont {Damour}\ and\ \citenamefont {Nagar}(2009)}]{Damour:2009vw}%
  \BibitemOpen
  \bibfield  {author} {\bibinfo {author} {\bibfnamefont {T.}~\bibnamefont {Damour}}\ and\ \bibinfo {author} {\bibfnamefont {A.}~\bibnamefont {Nagar}},\ }\bibfield  {title} {\bibinfo {title} {{Relativistic tidal properties of neutron stars}},\ }\href {https://doi.org/10.1103/PhysRevD.80.084035} {\bibfield  {journal} {\bibinfo  {journal} {Phys. Rev. D}\ }\textbf {\bibinfo {volume} {80}},\ \bibinfo {pages} {084035} (\bibinfo {year} {2009})},\ \Eprint {https://arxiv.org/abs/0906.0096} {arXiv:0906.0096 [gr-qc]} \BibitemShut {NoStop}%
\bibitem [{\citenamefont {Binnington}\ and\ \citenamefont {Poisson}(2009)}]{Binnington:2009bb}%
  \BibitemOpen
  \bibfield  {author} {\bibinfo {author} {\bibfnamefont {T.}~\bibnamefont {Binnington}}\ and\ \bibinfo {author} {\bibfnamefont {E.}~\bibnamefont {Poisson}},\ }\bibfield  {title} {\bibinfo {title} {{Relativistic theory of tidal Love numbers}},\ }\href {https://doi.org/10.1103/PhysRevD.80.084018} {\bibfield  {journal} {\bibinfo  {journal} {Phys. Rev. D}\ }\textbf {\bibinfo {volume} {80}},\ \bibinfo {pages} {084018} (\bibinfo {year} {2009})},\ \Eprint {https://arxiv.org/abs/0906.1366} {arXiv:0906.1366 [gr-qc]} \BibitemShut {NoStop}%
\bibitem [{\citenamefont {Poisson}(2021)}]{Poisson:2020vap}%
  \BibitemOpen
  \bibfield  {author} {\bibinfo {author} {\bibfnamefont {E.}~\bibnamefont {Poisson}},\ }\bibfield  {title} {\bibinfo {title} {{Compact body in a tidal environment: New types of relativistic Love numbers, and a post-Newtonian operational definition for tidally induced multipole moments}},\ }\href {https://doi.org/10.1103/PhysRevD.103.064023} {\bibfield  {journal} {\bibinfo  {journal} {Phys. Rev. D}\ }\textbf {\bibinfo {volume} {103}},\ \bibinfo {pages} {064023} (\bibinfo {year} {2021})},\ \Eprint {https://arxiv.org/abs/2012.10184} {arXiv:2012.10184 [gr-qc]} \BibitemShut {NoStop}%
\bibitem [{\citenamefont {Hegade K.~R.}\ \emph {et~al.}(2024)\citenamefont {Hegade K.~R.}, \citenamefont {Ripley},\ and\ \citenamefont {Yunes}}]{HegadeKR:2024agt}%
  \BibitemOpen
  \bibfield  {author} {\bibinfo {author} {\bibfnamefont {A.}~\bibnamefont {Hegade K.~R.}}, \bibinfo {author} {\bibfnamefont {J.~L.}\ \bibnamefont {Ripley}},\ and\ \bibinfo {author} {\bibfnamefont {N.}~\bibnamefont {Yunes}},\ }\bibfield  {title} {\bibinfo {title} {{Dynamical tidal response of nonrotating relativistic stars}},\ }\href {https://doi.org/10.1103/PhysRevD.109.104064} {\bibfield  {journal} {\bibinfo  {journal} {Phys. Rev. D}\ }\textbf {\bibinfo {volume} {109}},\ \bibinfo {pages} {104064} (\bibinfo {year} {2024})},\ \Eprint {https://arxiv.org/abs/2403.03254} {arXiv:2403.03254 [gr-qc]} \BibitemShut {NoStop}%
\bibitem [{\citenamefont {Riegert}(1984)}]{Riegert:1984kt}%
  \BibitemOpen
  \bibfield  {author} {\bibinfo {author} {\bibfnamefont {R.~J.}\ \bibnamefont {Riegert}},\ }\bibfield  {title} {\bibinfo {title} {{A Nonlocal Action for the Trace Anomaly}},\ }\href {https://doi.org/10.1016/0370-2693(84)90983-3} {\bibfield  {journal} {\bibinfo  {journal} {Phys. Lett. B}\ }\textbf {\bibinfo {volume} {134}},\ \bibinfo {pages} {56} (\bibinfo {year} {1984})}\BibitemShut {NoStop}%
\bibitem [{\citenamefont {Shapiro}\ and\ \citenamefont {Zheksenaev}(1994)}]{Shapiro:1994ww}%
  \BibitemOpen
  \bibfield  {author} {\bibinfo {author} {\bibfnamefont {I.~L.}\ \bibnamefont {Shapiro}}\ and\ \bibinfo {author} {\bibfnamefont {A.~G.}\ \bibnamefont {Zheksenaev}},\ }\bibfield  {title} {\bibinfo {title} {{Gauge dependence in higher derivative quantum gravity and the conformal anomaly problem}},\ }\href {https://doi.org/10.1016/0370-2693(94)90195-3} {\bibfield  {journal} {\bibinfo  {journal} {Phys. Lett. B}\ }\textbf {\bibinfo {volume} {324}},\ \bibinfo {pages} {286} (\bibinfo {year} {1994})}\BibitemShut {NoStop}%
\bibitem [{\citenamefont {Balbinot}\ \emph {et~al.}(1999{\natexlab{a}})\citenamefont {Balbinot}, \citenamefont {Fabbri},\ and\ \citenamefont {Shapiro}}]{Balbinot:1999ri}%
  \BibitemOpen
  \bibfield  {author} {\bibinfo {author} {\bibfnamefont {R.}~\bibnamefont {Balbinot}}, \bibinfo {author} {\bibfnamefont {A.}~\bibnamefont {Fabbri}},\ and\ \bibinfo {author} {\bibfnamefont {I.~L.}\ \bibnamefont {Shapiro}},\ }\bibfield  {title} {\bibinfo {title} {{Anomaly induced effective actions and Hawking radiation}},\ }\href {https://doi.org/10.1103/PhysRevLett.83.1494} {\bibfield  {journal} {\bibinfo  {journal} {Phys. Rev. Lett.}\ }\textbf {\bibinfo {volume} {83}},\ \bibinfo {pages} {1494} (\bibinfo {year} {1999}{\natexlab{a}})},\ \Eprint {https://arxiv.org/abs/hep-th/9904074} {arXiv:hep-th/9904074} \BibitemShut {NoStop}%
\bibitem [{\citenamefont {Balbinot}\ \emph {et~al.}(1999{\natexlab{b}})\citenamefont {Balbinot}, \citenamefont {Fabbri},\ and\ \citenamefont {Shapiro}}]{Balbinot:1999vg}%
  \BibitemOpen
  \bibfield  {author} {\bibinfo {author} {\bibfnamefont {R.}~\bibnamefont {Balbinot}}, \bibinfo {author} {\bibfnamefont {A.}~\bibnamefont {Fabbri}},\ and\ \bibinfo {author} {\bibfnamefont {I.~L.}\ \bibnamefont {Shapiro}},\ }\bibfield  {title} {\bibinfo {title} {{Vacuum polarization in Schwarzschild space-time by anomaly induced effective actions}},\ }\href {https://doi.org/10.1016/S0550-3213(99)00424-1} {\bibfield  {journal} {\bibinfo  {journal} {Nucl. Phys. B}\ }\textbf {\bibinfo {volume} {559}},\ \bibinfo {pages} {301} (\bibinfo {year} {1999}{\natexlab{b}})},\ \Eprint {https://arxiv.org/abs/hep-th/9904162} {arXiv:hep-th/9904162} \BibitemShut {NoStop}%
\bibitem [{\citenamefont {Mazur}\ and\ \citenamefont {Mottola}(2001)}]{Mazur:2001aa}%
  \BibitemOpen
  \bibfield  {author} {\bibinfo {author} {\bibfnamefont {P.~O.}\ \bibnamefont {Mazur}}\ and\ \bibinfo {author} {\bibfnamefont {E.}~\bibnamefont {Mottola}},\ }\bibfield  {title} {\bibinfo {title} {{Weyl cohomology and the effective action for conformal anomalies}},\ }\href {https://doi.org/10.1103/PhysRevD.64.104022} {\bibfield  {journal} {\bibinfo  {journal} {Phys. Rev. D}\ }\textbf {\bibinfo {volume} {64}},\ \bibinfo {pages} {104022} (\bibinfo {year} {2001})},\ \Eprint {https://arxiv.org/abs/hep-th/0106151} {arXiv:hep-th/0106151} \BibitemShut {NoStop}%
\bibitem [{\citenamefont {Mottola}\ and\ \citenamefont {Vaulin}(2006)}]{Mottola:2006ew}%
  \BibitemOpen
  \bibfield  {author} {\bibinfo {author} {\bibfnamefont {E.}~\bibnamefont {Mottola}}\ and\ \bibinfo {author} {\bibfnamefont {R.}~\bibnamefont {Vaulin}},\ }\bibfield  {title} {\bibinfo {title} {{Macroscopic Effects of the Quantum Trace Anomaly}},\ }\href {https://doi.org/10.1103/PhysRevD.74.064004} {\bibfield  {journal} {\bibinfo  {journal} {Phys. Rev. D}\ }\textbf {\bibinfo {volume} {74}},\ \bibinfo {pages} {064004} (\bibinfo {year} {2006})},\ \Eprint {https://arxiv.org/abs/gr-qc/0604051} {arXiv:gr-qc/0604051} \BibitemShut {NoStop}%
\bibitem [{\citenamefont {Barcelo}\ \emph {et~al.}(2012)\citenamefont {Barcelo}, \citenamefont {Carballo},\ and\ \citenamefont {Garay}}]{Barcelo:2011bb}%
  \BibitemOpen
  \bibfield  {author} {\bibinfo {author} {\bibfnamefont {C.}~\bibnamefont {Barcelo}}, \bibinfo {author} {\bibfnamefont {R.}~\bibnamefont {Carballo}},\ and\ \bibinfo {author} {\bibfnamefont {L.~J.}\ \bibnamefont {Garay}},\ }\bibfield  {title} {\bibinfo {title} {{Two formalisms, one renormalized stress-energy tensor}},\ }\href {https://doi.org/10.1103/PhysRevD.85.084001} {\bibfield  {journal} {\bibinfo  {journal} {Phys. Rev. D}\ }\textbf {\bibinfo {volume} {85}},\ \bibinfo {pages} {084001} (\bibinfo {year} {2012})},\ \Eprint {https://arxiv.org/abs/1112.0489} {arXiv:1112.0489 [gr-qc]} \BibitemShut {NoStop}%
\bibitem [{\citenamefont {Shen}\ \emph {et~al.}(2015)\citenamefont {Shen}, \citenamefont {Izumi},\ and\ \citenamefont {Chen}}]{Shen:2015zya}%
  \BibitemOpen
  \bibfield  {author} {\bibinfo {author} {\bibfnamefont {C.-M.}\ \bibnamefont {Shen}}, \bibinfo {author} {\bibfnamefont {K.}~\bibnamefont {Izumi}},\ and\ \bibinfo {author} {\bibfnamefont {P.}~\bibnamefont {Chen}},\ }\bibfield  {title} {\bibinfo {title} {{Boundary effect of anomaly-induced action}},\ }\href {https://doi.org/10.1103/PhysRevD.92.024035} {\bibfield  {journal} {\bibinfo  {journal} {Phys. Rev. D}\ }\textbf {\bibinfo {volume} {92}},\ \bibinfo {pages} {024035} (\bibinfo {year} {2015})},\ \bibinfo {note} {[Addendum: Phys.Rev.D 92, 049902 (2015)]},\ \Eprint {https://arxiv.org/abs/1505.00959} {arXiv:1505.00959 [gr-qc]} \BibitemShut {NoStop}%
\bibitem [{\citenamefont {Carroll}(2019)}]{Carroll:2004st}%
  \BibitemOpen
  \bibfield  {author} {\bibinfo {author} {\bibfnamefont {S.~M.}\ \bibnamefont {Carroll}},\ }\href {https://doi.org/10.1017/9781108770385} {\emph {\bibinfo {title} {{Spacetime and Geometry}: {An Introduction to General Relativity}}}}\ (\bibinfo  {publisher} {Cambridge University Press},\ \bibinfo {year} {2019})\BibitemShut {NoStop}%
\bibitem [{\citenamefont {Knizhnik}\ \emph {et~al.}(1988)\citenamefont {Knizhnik}, \citenamefont {Polyakov},\ and\ \citenamefont {Zamolodchikov}}]{Knizhnik:1988ak}%
  \BibitemOpen
  \bibfield  {author} {\bibinfo {author} {\bibfnamefont {V.~G.}\ \bibnamefont {Knizhnik}}, \bibinfo {author} {\bibfnamefont {A.~M.}\ \bibnamefont {Polyakov}},\ and\ \bibinfo {author} {\bibfnamefont {A.~B.}\ \bibnamefont {Zamolodchikov}},\ }\bibfield  {title} {\bibinfo {title} {{Fractal Structure of 2D Quantum Gravity}},\ }\href {https://doi.org/10.1142/S0217732388000982} {\bibfield  {journal} {\bibinfo  {journal} {Mod. Phys. Lett. A}\ }\textbf {\bibinfo {volume} {3}},\ \bibinfo {pages} {819} (\bibinfo {year} {1988})}\BibitemShut {NoStop}%
\bibitem [{\citenamefont {David}(1988)}]{David:1988hj}%
  \BibitemOpen
  \bibfield  {author} {\bibinfo {author} {\bibfnamefont {F.}~\bibnamefont {David}},\ }\bibfield  {title} {\bibinfo {title} {{Conformal Field Theories Coupled to 2D Gravity in the Conformal Gauge}},\ }\href {https://doi.org/10.1142/S0217732388001975} {\bibfield  {journal} {\bibinfo  {journal} {Mod. Phys. Lett. A}\ }\textbf {\bibinfo {volume} {3}},\ \bibinfo {pages} {1651} (\bibinfo {year} {1988})}\BibitemShut {NoStop}%
\bibitem [{\citenamefont {Distler}\ and\ \citenamefont {Kawai}(1989)}]{Distler:1988jt}%
  \BibitemOpen
  \bibfield  {author} {\bibinfo {author} {\bibfnamefont {J.}~\bibnamefont {Distler}}\ and\ \bibinfo {author} {\bibfnamefont {H.}~\bibnamefont {Kawai}},\ }\bibfield  {title} {\bibinfo {title} {{Conformal Field Theory and 2D Quantum Gravity}},\ }\href {https://doi.org/10.1016/0550-3213(89)90354-4} {\bibfield  {journal} {\bibinfo  {journal} {Nucl. Phys. B}\ }\textbf {\bibinfo {volume} {321}},\ \bibinfo {pages} {509} (\bibinfo {year} {1989})}\BibitemShut {NoStop}%
\bibitem [{\citenamefont {Antoniadis}\ \emph {et~al.}(1997)\citenamefont {Antoniadis}, \citenamefont {Mazur},\ and\ \citenamefont {Mottola}}]{Antoniadis:1995dy}%
  \BibitemOpen
  \bibfield  {author} {\bibinfo {author} {\bibfnamefont {I.}~\bibnamefont {Antoniadis}}, \bibinfo {author} {\bibfnamefont {P.~O.}\ \bibnamefont {Mazur}},\ and\ \bibinfo {author} {\bibfnamefont {E.}~\bibnamefont {Mottola}},\ }\bibfield  {title} {\bibinfo {title} {{Physical states of the quantum conformal factor}},\ }\href {https://doi.org/10.1103/PhysRevD.55.4770} {\bibfield  {journal} {\bibinfo  {journal} {Phys. Rev. D}\ }\textbf {\bibinfo {volume} {55}},\ \bibinfo {pages} {4770} (\bibinfo {year} {1997})},\ \Eprint {https://arxiv.org/abs/hep-th/9509169} {arXiv:hep-th/9509169} \BibitemShut {NoStop}%
\bibitem [{\citenamefont {Brown}\ and\ \citenamefont {Ottewill}(1986)}]{Brown:1986tj}%
  \BibitemOpen
  \bibfield  {author} {\bibinfo {author} {\bibfnamefont {M.~R.}\ \bibnamefont {Brown}}\ and\ \bibinfo {author} {\bibfnamefont {A.~C.}\ \bibnamefont {Ottewill}},\ }\bibfield  {title} {\bibinfo {title} {{Photon Propagators and the Definition and Approximation of Renormalized Stress Tensors in Curved Space-time}},\ }\href {https://doi.org/10.1103/PhysRevD.34.1776} {\bibfield  {journal} {\bibinfo  {journal} {Phys. Rev. D}\ }\textbf {\bibinfo {volume} {34}},\ \bibinfo {pages} {1776} (\bibinfo {year} {1986})}\BibitemShut {NoStop}%
\bibitem [{\citenamefont {Hayward}(2006)}]{Hayward:2005gi}%
  \BibitemOpen
  \bibfield  {author} {\bibinfo {author} {\bibfnamefont {S.~A.}\ \bibnamefont {Hayward}},\ }\bibfield  {title} {\bibinfo {title} {{Formation and evaporation of regular black holes}},\ }\href {https://doi.org/10.1103/PhysRevLett.96.031103} {\bibfield  {journal} {\bibinfo  {journal} {Phys. Rev. Lett.}\ }\textbf {\bibinfo {volume} {96}},\ \bibinfo {pages} {031103} (\bibinfo {year} {2006})},\ \Eprint {https://arxiv.org/abs/gr-qc/0506126} {arXiv:gr-qc/0506126} \BibitemShut {NoStop}%
\bibitem [{\citenamefont {Inc.}(2023)}]{Mathematica}%
  \BibitemOpen
  \bibfield  {author} {\bibinfo {author} {\bibfnamefont {W.~R.}\ \bibnamefont {Inc.}},\ }\href {https://www.wolfram.com/mathematica} {\bibinfo {title} {Mathematica, {V}ersion 13.3}} (\bibinfo {year} {2023}),\ \bibinfo {note} {champaign, IL, 2023}\BibitemShut {NoStop}%
\bibitem [{\citenamefont {Dvali}\ \emph {et~al.}(2009)\citenamefont {Dvali}, \citenamefont {Sawicki},\ and\ \citenamefont {Vikman}}]{Dvali:2009fw}%
  \BibitemOpen
  \bibfield  {author} {\bibinfo {author} {\bibfnamefont {G.}~\bibnamefont {Dvali}}, \bibinfo {author} {\bibfnamefont {I.}~\bibnamefont {Sawicki}},\ and\ \bibinfo {author} {\bibfnamefont {A.}~\bibnamefont {Vikman}},\ }\bibfield  {title} {\bibinfo {title} {{Dark Matter via Many Copies of the Standard Model}},\ }\href {https://doi.org/10.1088/1475-7516/2009/08/009} {\bibfield  {journal} {\bibinfo  {journal} {JCAP}\ }\textbf {\bibinfo {volume} {08}},\ \bibinfo {pages} {009}},\ \Eprint {https://arxiv.org/abs/0903.0660} {arXiv:0903.0660 [hep-th]} \BibitemShut {NoStop}%
\bibitem [{\citenamefont {Dvali}\ and\ \citenamefont {Redi}(2009)}]{Dvali:2009ne}%
  \BibitemOpen
  \bibfield  {author} {\bibinfo {author} {\bibfnamefont {G.}~\bibnamefont {Dvali}}\ and\ \bibinfo {author} {\bibfnamefont {M.}~\bibnamefont {Redi}},\ }\bibfield  {title} {\bibinfo {title} {{Phenomenology of $10^{32}$ Dark Sectors}},\ }\href {https://doi.org/10.1103/PhysRevD.80.055001} {\bibfield  {journal} {\bibinfo  {journal} {Phys. Rev. D}\ }\textbf {\bibinfo {volume} {80}},\ \bibinfo {pages} {055001} (\bibinfo {year} {2009})},\ \Eprint {https://arxiv.org/abs/0905.1709} {arXiv:0905.1709 [hep-ph]} \BibitemShut {NoStop}%
\bibitem [{\citenamefont {Davies}\ \emph {et~al.}(1976)\citenamefont {Davies}, \citenamefont {Fulling},\ and\ \citenamefont {Unruh}}]{Davies:1976ei}%
  \BibitemOpen
  \bibfield  {author} {\bibinfo {author} {\bibfnamefont {P.~C.~W.}\ \bibnamefont {Davies}}, \bibinfo {author} {\bibfnamefont {S.~A.}\ \bibnamefont {Fulling}},\ and\ \bibinfo {author} {\bibfnamefont {W.~G.}\ \bibnamefont {Unruh}},\ }\bibfield  {title} {\bibinfo {title} {{Energy Momentum Tensor Near an Evaporating Black Hole}},\ }\href {https://doi.org/10.1103/PhysRevD.13.2720} {\bibfield  {journal} {\bibinfo  {journal} {Phys. Rev. D}\ }\textbf {\bibinfo {volume} {13}},\ \bibinfo {pages} {2720} (\bibinfo {year} {1976})}\BibitemShut {NoStop}%
\bibitem [{\citenamefont {Parentani}\ and\ \citenamefont {Piran}(1994)}]{Parentani:1994ij}%
  \BibitemOpen
  \bibfield  {author} {\bibinfo {author} {\bibfnamefont {R.}~\bibnamefont {Parentani}}\ and\ \bibinfo {author} {\bibfnamefont {T.}~\bibnamefont {Piran}},\ }\bibfield  {title} {\bibinfo {title} {{The Internal geometry of an evaporating black hole}},\ }\href {https://doi.org/10.1103/PhysRevLett.73.2805} {\bibfield  {journal} {\bibinfo  {journal} {Phys. Rev. Lett.}\ }\textbf {\bibinfo {volume} {73}},\ \bibinfo {pages} {2805} (\bibinfo {year} {1994})},\ \Eprint {https://arxiv.org/abs/hep-th/9405007} {arXiv:hep-th/9405007} \BibitemShut {NoStop}%
\bibitem [{\citenamefont {Mottola}\ \emph {et~al.}(2023)\citenamefont {Mottola}, \citenamefont {Chandra}, \citenamefont {Manca},\ and\ \citenamefont {Sorkin}}]{Mottola:2023jlo}%
  \BibitemOpen
  \bibfield  {author} {\bibinfo {author} {\bibfnamefont {E.}~\bibnamefont {Mottola}}, \bibinfo {author} {\bibfnamefont {M.}~\bibnamefont {Chandra}}, \bibinfo {author} {\bibfnamefont {G.~M.}\ \bibnamefont {Manca}},\ and\ \bibinfo {author} {\bibfnamefont {E.}~\bibnamefont {Sorkin}},\ }\bibfield  {title} {\bibinfo {title} {{Quantum effects of the conformal anomaly in a 2D model of gravitational collapse}},\ }\href {https://doi.org/10.1007/JHEP08(2023)223} {\bibfield  {journal} {\bibinfo  {journal} {JHEP}\ }\textbf {\bibinfo {volume} {08}},\ \bibinfo {pages} {223}},\ \Eprint {https://arxiv.org/abs/2303.15397} {arXiv:2303.15397 [gr-qc]} \BibitemShut {NoStop}%
\end{thebibliography}%

\end{document}